\documentclass[amsmath,amssymb,aps,reprint]{revtex4-2}
\usepackage{xcolor}
\usepackage{graphicx}
\usepackage{bm,graphicx}
\usepackage{hyperref}
\usepackage{subcaption}
\usepackage{pgfplots}
\usepackage{xparse}
\usepackage{feynmf}
\usepackage{tikz-feynman}
\tikzfeynmanset{compat=1.1.0}
\usepackage{amsmath}
\usepackage{diagbox}
\begin{document}

\title{Double Higgs Boson Production via Photon Fusion at Muon Colliders within the Triplet Higgs Model }

\author{Bathiya Samarakoon}
\email{bxsamarakoonmudeligedon@shockers.wichita.edu}
\author{ Terrance M. Figy}
\email{terrance.figy@wichita.edu}
\affiliation{Department of Mathematics, Statistics and Physics, Wichita State University
1845, Fairmount St, Wichita, KS, 67260.
}

\date{\today}

\begin{abstract}
  In this paper, we present predictions for scattering cross section the of Higgs boson pair production via photon fusion at future muon colliders, focusing specifically the production processes $\mu^+\mu^- \rightarrow \gamma\gamma \rightarrow h^{0}h^{0}, A^{0}A^{0}$. We investigated the impact of three choices the photon structure functions on cross-section predictions for a range model input parameters within the theoretical framework of the Higgs triplet model\cite{Arhrib:2011uy,Rahili:2019ixf,PhysRevD.22.2227,PhysRevD.25.2951}.
\end{abstract}
\maketitle
\section{Introduction}
\label{sec:introduction}

Double Higgs production is important for testing the Higgs self-coupling \cite{Grober:2016yuo, ATLAS:2012yve} which is responsible for providing mass to elementary particles and the shape of the Higgs potential. In the SM, the Higgs potential for the Higgs field $\phi$ is defined as,
\begin{equation}\label{eq:potentialsm}
    V(\phi)=\mu^2\phi\phi^* + \frac{\lambda}{2}(\phi^*\phi)^2 \,
\end{equation}
where $\lambda >0$ and $\mu^2<0$. The minimum value of the Higgs potential occurs at $v^2=-\mu^2/\lambda$. 
Following spontaneous symmetry breaking, the Higgs boson acquires a mass $M_{H}=-2\mu^2=2\lambda v^2$. In the Standard Model, the relationships among the physical Higgs mass, cubic interaction ($\lambda_{hhh}$), and quartic interaction ($\lambda_{hhhh}$) are uniquely defined and can be expressed as $\lambda_{hhh}=v\lambda_{hhhh}=3M_{H}^2/v$.
However, the trilinear Higgs coupling is challenging to measure directly, as it requires the production of two or more Higgs bosons simultaneously \cite{Spira:2016ztx, Asakawa:2008se, Baglio:2020ini}. Also, to measure the trilinear Higgs coupling at the LHC requires high luminosity because processes that involve the trilinear Higgs coupling are rare in SM \cite{Asakawa:2008se}. The anticipated cross section for the production of Higgs boson pairs via gluon gluon fusion stands at approximately $36.69 \ \mathrm{fb}$ at a center of mass energy of 14 TeV~\cite{ATL-PHYS-PUB-2018-053,Cepeda:2019klc}. Even with the highest possible enhancement of both center of mass energy and integrated luminosity at LHC, the accurate extraction of $\lambda_{hhhh}$ remains a formidable challenge. Nevertheless, there is optimism regarding the feasible observation of Higgs boson pair production and the determination of $\lambda_{hhh}$. However, achieving these goals might necessitate an integrated luminosity of $3000 \ \mathrm{fb}^{-1}$ at the HL-LHC, \cite{ATL-PHYS-PUB-2018-053,Chiesa:2020awd,Portales:2022ggy}. The production of Higgs boson pairs via gluon fusion has been studied across various theoretical frameworks, including the Standard Model \cite{Glover:1987nx}, minimal supersymmetric Standard Model \cite{Dawson:1998py,Plehn:1996wb}, Lee Wick Standard Model \cite{Figy:2011yu}, and singlet extension model \cite{Dawson:2015haa}.

Muon colliders offer several advantages over proton colliders, which can help to avoid some of the challenges associated with measuring the trilinear Higgs coupling. Muon colliders can reach higher center of mass energies than proton colliders [$\sqrt{s} \sim \mathcal O (10 \ \mathrm{ TeV}) $], which can increase the production rate of triple Higgs boson events\cite{Chiesa:2021qpr}. Here we want to analyze the double Higgs production initiated by collinear photons radiated by high energy muon beams. 
The spectrum of photons with an energy fraction 
$x$ emitted by a charged lepton with an initial energy $E$ is described by the Weizsäcker-Williams spectrum, also known as the leading order effective photon approximation (EPA) \cite{Delahaye:2019omf,Schwartz:2014sze},
 \begin{equation}\label{eq:EPALO}
     f_{\gamma,l}(x) \approx \frac{\alpha_e}{2\pi}P_{\gamma,l}(x)\ln{\frac{E^2}{m_l^2}} \ .
 \end{equation}
The splitting functions are,
$ P_{\gamma,l}=(1+(1-x^2))/x$ for $l\rightarrow{\gamma}$, and $P_{l/l}(x)=(1+x^2)/(1-x)$ for $l\rightarrow{l}$. 

Double Higgs boson production via photon fusion has been examined in the SM at multi-TeV muon collider \cite{Chiesa:2021qpr} and in the Two Higgs doublet model (2HDM) at the International Linear Collider (ILC) \cite{Arhrib:2009gg, Demirci:2020zgt, Demirci:2019kop}. However, these models do not account for the impact of the $H^{\pm\pm}$ boson's involvement in the $\gamma\gamma \rightarrow h^0h^0, A^0A^0$ processes. This aspect is addressed within the context of the Higgs triplet model (HTM). In the HTM, a noteworthy hierarchy is observed between the masses of $H^{\pm\pm}$ and $H^{\pm}$, and this influence is thoroughly investigated in the context of this paper.

Initially, we examine the partonic cross sections involving $\gamma\gamma$ fusion leading to the production of Higgs pairs in the HTM. Following that, we numerically compute the cross sections for $\mu^+\mu^- \rightarrow \gamma\gamma \rightarrow h^0h^0,A^0A^0$ by performing the convolution of the parton distribution functions (PDFs) with the partonic scattering cross section. We consider the muon collider at a benchmark energy of $\sqrt{s} = 3$ TeV, with the integrated luminosity \cite{Han:2022edd} scaled according to
\begin{align*}\label{lumi}
\mathcal{L} = \left(\frac{\sqrt{s}}{10 \ \mathrm{TeV}}\right)^2 \times 10^{4} \ \mathrm{{fb}^{-1}} \approx 1 \ \mathrm{ab^{-1}}.
\end{align*}
This paper is organized as follows: In Sec.~\ref{sec:theory}, we present a description of the Higgs triplet model, while Sec.~\ref{sec:potential} provides constraints on the model parameters and input values. Sec.~\ref{sec:method} outlines the computational method used for calculating the scattering amplitude and cross sections. Moving to Sec.~\ref{sec:results}, we conduct an analysis and present numerical results encompassing partonic cross sections, collider simulations for $h^0h^0$ and $A^0A^0$ production, along with the discussion. Additionally, we explore the examined couplings, particularly addressing the decoupling and weak-coupling limits that may arise in the calculations. We summarize our findings in Sec.~\ref{sec:conclusions}.

\section{Theoretical Framework}
\label{sec:theory}

Our theoretical framework is built upon the HTM, also known as the type II seesaw \cite{Arhrib:2011uy,Rahili:2019ixf, PhysRevD.22.2227,PhysRevD.25.2951}. In the HTM, alongside the Standard Model weak doublet, represented as $\Phi \sim (1,2,1/2)$, there is an additional Higgs triplet denoted as $\Delta \sim (1,3,2)$ which transforms under the $SU(2)_L$ gauge group.
 The motivation for the type II seesaw Model stems from the observation that two doublets can be decomposed into a triplet and a singlet representation  ($2 \otimes 2 = 3 \oplus 1$). It is assumed that these additional fields possess a mass scale that is significantly higher than the electroweak scale. By introducing these extra scalar fields, new Yukawa couplings are established between the SM lepton fields and the scalar fields. These Yukawa couplings generate relatively small Majorana neutrino masses without requiring right handed neutrinos. Due to the higher mass scale of the additional scalar fields, the resulting masses for the neutrinos are much smaller compared to the electroweak scale. In addition to the Yukawa interactions present in the Standard Model, the Yukawa sector of the type II seesaw model incorporates interactions between the Higgs triplet field ($\Delta$) and the lepton fields:
\begin{equation}\label{eq:leptonfield}
    \mathcal{L}_{Y} \supset Y_{\nu}L^T C \otimes i \sigma^2 \Delta L + h.c.
\end{equation}
where $L$ represents a left-handed lepton doublet, $C$ denotes the Dirac charge conjugation operator, $\sigma_a$ (with $a=1,2,3$) refers to the Pauli matrices, and $Y_{\nu}$ represents the Yukawa couplings for neutrinos. 

On the other hand, the kinetic and gauge interactions of the new field $\Delta$ are embodied in the Lagrangian term, which takes the following form:
\begin{equation}\label{eq:kineticterm} \mathcal{L}_{k}=Tr[(D_{\mu}\Delta)^{\dagger}(D^{\mu}\Delta)],
\end{equation}
incorporating the covariant derivative $D_{\mu}\Delta = \partial_{\mu}\Delta + i\frac{g}{2}[\sigma^aW^a_{\mu},\Delta]+i\frac{g^{'}}{2}Y_{\Delta}B_{\mu}\Delta$.

The general scalar potential term ,$V(\Phi,\Delta)$, which is renormalizable, \textit{CP}-invariant, and gauge invariant, can be expressed as follows:
\begin{eqnarray} \label{eq:potentailterm}
V(\Phi,\Delta)&=&-\mu_{\Phi}^2\Phi^{\dagger}\Phi + \frac{\lambda}{4}(\Phi^{\dagger}\Phi)^2 + \mu^2_{\Delta}Tr(\Delta^{\dagger}\Delta) \\
&+&[\mu(\Phi^T i\sigma^2 \Delta^{\dagger} \Phi)+ h.c]  + \lambda_{1}(\Phi^{\dagger}\Phi) Tr(\Delta^{\dagger}\Delta) \nonumber \\
&+& \lambda_{2} (Tr \Delta^{\dagger}\Delta)^{2} 
 + \lambda_{3} Tr(\Delta^{\dagger}\Delta)^{2} 
 + \lambda_{4} \Phi^{\dagger} \Delta \Delta^{\dagger} \Phi \nonumber \ . 
\end{eqnarray}

The doublet field $\Phi$, characterized by a weak hypercharge $Y_{\Phi} = 1$, and the triplet field $\Delta$, which is represented in a $2\times2$ representation with a weak hypercharge $Y_{\Delta} = 2$, are expressed as,

\begin{equation*}\label{eq:fields}
    \Phi = \begin{pmatrix}
        \phi^+  \\
        \phi^0 
    \end{pmatrix}, 
    \quad 
    \Delta = \begin{pmatrix}
        \frac{\delta^{+}}{\sqrt{2}} & \delta^{++} \\
\delta^0 & -\frac{\delta^+}{\sqrt{2}}   
    \end{pmatrix}  \ . 
\end{equation*}

The Higgs triplet model incorporates five dimensionless parameters ($\lambda$, $\lambda_1$, $\lambda_2$, $\lambda_3$, $\lambda_4$), two real parameters with mass dimensions ($\mu_\Phi$, $\mu_\Delta$), and a lepton number violating parameter with positive mass dimension ($\mu$). The vacuum expectation value ($v_\Phi$) of the Higgs field, which is responsible for breaking the electroweak symmetry, takes place in a direction that does not introduce any electric charge or alter the electrically neutral nature of the vacuum. The Higgs triplet field acquires a nonzero vacuum expectation value ($v_\Delta$) that leads to the spontaneous breaking of the electroweak symmetry. Consequently, the neutral components of the Higgs triplet field develop nonzero vacuum expectation value, while the charged components remain zero.

In the absence of \textit{CP} violation, the scalar fields $\phi^0$ and $\delta^0$ can be parametrized as follows: $\phi^0 = \frac{1}{\sqrt{2}}(v_\Phi+\phi+i\chi)$ and $\delta^0 = \frac{1}{\sqrt{2}}(v_\Delta+\delta+i\eta)$. This parameterization involves shifting the real part of $\phi^0$, denoted as $\phi$, and the real part of $\delta^0$, denoted as $\delta$, around their vacuum expectation values (VEVs). As a result, a $10\times10$ squared mass matrix is obtained to describe the scalars in the model. After diagonalizing the mass matrix and utilizing the following rotation matrices:
\begin{equation}\label{rot1}
\begin{pmatrix}
h^0 \\ H^0 \\ 
\end{pmatrix}
=
\begin{pmatrix}
\cos{\alpha} & \sin{\alpha} \\
-\sin{\alpha} & \cos{\alpha} \\
\end{pmatrix}
\begin{pmatrix}
\phi \\ \delta
\end{pmatrix}
\end{equation},

\begin{equation}\label{rot2}
\begin{pmatrix}
G^{\pm} \\ H^{\pm} \\ 
\end{pmatrix}
=
\begin{pmatrix}
\cos{\beta^{'}} & \sin{\beta^{'}} \\
-\sin{\beta^{'}} & \cos{\beta^{'}} \\
\end{pmatrix}
\begin{pmatrix}
\phi^{\pm} \\ \delta^{\pm}
\end{pmatrix}
\end{equation} and, 

\begin{equation}\label{rot3}
\begin{pmatrix}
G^{0} \\ A^{0} \\ 
\end{pmatrix}
=
\begin{pmatrix}
\cos{\beta} & \sin{\beta} \\
-\sin{\beta} & \cos{\beta} \\
\end{pmatrix}
\begin{pmatrix}
\chi \\ \eta
\end{pmatrix}
\end{equation}
to transform the fields into their mass eigenstates, a total of six physical Higgs states ($A^0,H^0,H^{\pm},H^{\pm\pm}$), in addition to the Standard Model Higgs boson ($h^0$), as well as three massless Goldstone bosons ($G^0,G^{\pm}$) which acquire the role of the longitudinal components of the $Z^0$ and $W^{\pm}$ bosons, emerge within the model. Here the mixing angles are given by $\tan{\beta}=\sqrt{2}\tan{\beta^{'}}=2v_\Delta/v_\Phi$. The physical masses of the doubly charged and singly charged Higgs boson are expressed as,

\begin{equation}\label{eq:massCharged}
\begin{aligned}
    & M^2_{H^{\pm\pm}}=M_{\Delta}^2-v_{\Delta}^2\lambda_3-\frac{\lambda_4}{2}v_{\Phi}^2, \\
    & M^2_{H^{\pm}} =\Big(M_{\Delta}^2-\frac{\lambda_4}{4}v_{\Phi}^2\Big) \Big(1+\frac{2v^2_{\Delta}}{v^2_{\Phi}} \Big) \ , \\
    & M_{\Delta}^2 = \frac{\mu v^2_{\Phi}}{\sqrt{2}{v_{\Delta}}} \ .
\end{aligned}
\end{equation}

The \textit{CP}-even Higgs bosons, which correspond to the mass eigenstates resulting from the mixing of the doublet scalar field ($\phi$) and the triplet scalar field ($\Delta$) as shown in the Eq.~(\ref{rot1}), have mass eigenvalues determined by the following expressions:
\begin{equation}\label{eq:massnuetral}
\begin{aligned}
    & M_{h^0}^2 = \mathcal{K}_{1}^2\cos^2{\alpha}+\mathcal{K}_{2}^2\sin^2{\alpha}-\mathcal{K}_{3}^2\sin^2{2\alpha} \ , \\
    & M_{H^0}^2 = \mathcal{K}_{1}^2\cos^2{\alpha}+\mathcal{K}_{2}^2\sin^2{\alpha}+\mathcal{K}_{3}^2\sin^2{2\alpha} \ .
\end{aligned}
\end{equation}
The following coefficients $\mathcal{K}_{1}$, $\mathcal{K}_{2}$, and $\mathcal{K}_{3}$ are defined as follows:
\begin{equation}\label{eq:parameterK}
\begin{aligned}
    & \mathcal{K}_1^2 = \frac{v_{\Delta}^2\lambda}{2} \ , \\
    & \mathcal{K}_2^2 = M_{\Delta}^2+2v^2_{\Delta}(\lambda_2+\lambda_3) \,  \\
    & \mathcal{K}^2_3 = -\frac{2v_{\Delta}}{v_{\Phi}}M^2_{\Delta}+v_{\Delta}v_{\Phi}(\lambda_1+\lambda_4) \ .
\end{aligned}
\end{equation}

Similarly, the emergence of the pseudo scalar $A^0$ is attributed to the mixing between the fields $\chi$ and $\eta$, and its mass is determined by the expression:

\begin{equation}\label{eq:masspesudo}
M_{A^0}^2 = M_\Delta^2 \left(1 + \frac{4v_{\Delta}^2}{v_{\Phi}^2} \right).
\end{equation}
By expressing the couplings ($\lambda$,$\lambda_{1,2,3,4}$ and $\mu$) in terms of physical Higgs masses, the mixing angle $\alpha$ and VEVs ($v_{\Delta}$ and $v_{\Phi}$), we can directly relate the strength of the interactions to the masses of the particles involved and the vacuum expectation values that characterize the symmetry breaking. Furthermore, through this consistent parametrization of the couplings, we can readily compare the predictions and implications of the HTM with those of other models and experimental observations. Now, using Eq.~(\ref{eq:massCharged}),~(\ref{eq:massnuetral}) and~(\ref{eq:parameterK}), one can obtain the following relations:
\begin{equation}\label{eq:mu}
    \begin{aligned}
        \mu = \frac{\sqrt{2}v_{\Phi}}{v_{\Phi}^2+4v^2_{\Delta}}M^2_A \ ,
    \end{aligned}
\end{equation} 

\begin{equation}\label{eq:lam}
    \begin{aligned}
        \lambda = -\frac{2}{v_{\Phi}^2}(c_{\alpha}^2M^2_{h}+s^2_{\alpha}M^2_H) \ ,
    \end{aligned}
\end{equation}

\begin{equation}\label{eq:lam1}
    \begin{aligned}
        \lambda_1 = -\frac{2}{v_{\Phi}^2+4v^2_{\Delta}}M_A^2+ \frac{2}{v_{\Phi}^2+2v^2_{\Delta}}M_{H^{\pm}}^2\\  
        + \frac{\sin{2\alpha}}{2v_{\Delta}v_{\Phi}}(M_h^2-M_H^2) \ ,
    \end{aligned}
\end{equation}
 
\begin{equation}\label{eq:lam2}
    \begin{aligned}
        \lambda_2 = \frac{1}{2v_{\Delta}^2} \Big( c_{\alpha}^2M^2_{h}+s^2_{\alpha}M^2_H + \frac{v^2_{\Delta}}{v_{\Phi}^2+4v^2_{\Delta}}M_A^2 - \\ \frac{4v^2_{\Delta}}{v_{\Phi}^2+2v^2_{\Delta}}M_A^2 +2M^2_{H^{\pm\pm}}\Big) \ , 
    \end{aligned}
\end{equation}

\begin{equation}\label{eq:lam3}
    \begin{aligned}
        \lambda_3 = \frac{1}{v_{\Delta}^2} \Big(  -\frac{v^2_{\Phi}}{v_{\Phi}^2+4v^2_{\Delta}}M_A^2 + \frac{2v^2_{\Phi}}{v_{\Phi}^2+2v^2_{\Delta}}M^2_{H^{\pm}} -M^2_{H^{\pm\pm}}\Big) 
    \end{aligned}
\end{equation} and,

\begin{equation}\label{eq:lam4}
    \begin{aligned}
        \lambda_4 =   \frac{4}{v_{\Phi}^2+4v^2_{\Delta}}M_A^2 - \frac{4}{v_{\Phi}^2+2v^2_{\Delta}}M^2_{H^{\pm}} \ .
    \end{aligned} 
\end{equation} 

Here we define two distinct parameter spaces as follows:
\begin{equation}
\begin{aligned}
\mathcal{P}_1= \{\mu, \lambda, \lambda_1, \lambda_2, \lambda_3, \lambda_4, \tan{\beta}, \cos{\alpha}\},
\end{aligned}
\end{equation}

\begin{equation}
\begin{aligned}
\mathcal{P}_2= \{ M_{h}, M_{H}, M_{A^0}, M_{H^{\pm}}, M_{H^{\pm\pm}}, v_{\Delta}, v_{\Phi}, \cos{\alpha}\}.
\end{aligned}
\end{equation}

By utilizing these parameter spaces, computations can be performed in a bidirectional manner, allowing for the evaluation of the quantities and relations within the HTM using either set of parameters. Moreover,  $v_{\Delta}$ and $ v_{\Phi}$ are reparametrized following the conventions of Ref.~\cite{Arhrib:2011uy}: 
\begin{equation}
    v_{\phi}^2 = \frac{1}{1+\frac{1}{2}\tan^2{\beta}}\frac{S_W^2M_W^2}{\pi\alpha_e} 
\end{equation}
and, 
\begin{equation}
    v_{\Delta}^2 = \frac{\tan^2{\beta}}{1+\frac{1}{2}\tan^2{\beta}}\frac{S_W^2M_W^2}{4\pi\alpha_e}
\end{equation}
with $v_{\phi}^2+2v_{\Delta}^2 = v^2$.

\section{Constraints on the potential}
\label{sec:potential}

Constraints on the potential in the HTM are necessary to ensure the model's consistency, stability, and compatibility with experimental observations. The potential plays a crucial role in determining the behavior of the scalar fields and their interactions. 

In the HTM, vacuum stability is required to ensure that the electroweak symmetry breaking minimum of the Higgs potential is stable and that the vacuum dies decay into a lower-energy state. In order to ensure that the scalar potential $V(\Phi,\Delta)$ in the HTM is always bounded from below (BFB) and does not lead to instability of the vacuum state, we must impose the following constraints on the model parameters from Ref.~\cite{Arhrib:2011uy,BhupalDev:2013xol}:
\begin{equation}
 \lambda \geq 0 ; \lambda_2+\lambda_3 \geq 0 ; \lambda_2+\frac{\lambda_3}{2} \geq 0 \ ,
 \label{eq:b1}
\end{equation}

\begin{equation}
    \lambda_1+\sqrt{\lambda(\lambda_2+\lambda_3)}\geq 0 ; \lambda_1+ 
\sqrt{\lambda(\lambda_2+\frac{\lambda_3}{2})} \geq 0 
\label{eq:b2}
\end{equation}
and,
\begin{equation}
    \lambda_1+\lambda_4+\sqrt{\lambda(\lambda_2+\lambda_3)}\geq 0 ; \lambda_1+\lambda_4+ 
\sqrt{\lambda(\lambda_2+\frac{\lambda_3}{2})} \geq 0 \ .
\label{eq:b3}
\end{equation}

To prevent the occurrence of tachyonic Higgs states, we establish the following constraints using Eq.~(\ref{eq:massCharged}) and Eq.~(\ref{eq:masspesudo}): 
\begin{equation}\label{mu1}
    \mu > \mu_{1} = 0 \ ,
\end{equation}
\begin{equation}\label{mu2}
    \mu > \mu_{2} = \frac{\lambda_4 M_W S_W}{4\sqrt{2\pi\alpha_e}} \frac{\tan{\beta}}{\sqrt{1+\frac{\tan^2{\beta}}{2}}} \ ,
\end{equation}
\begin{equation}\label{mu3}
    \mu > \mu_{3} = \frac{M_W S_W}{4\sqrt{2\pi\alpha_e}} \frac{(2\lambda_4\tan{\beta}+\lambda_3\tan^3{\beta})}{\sqrt{1+\frac{\tan^2{\beta}}{2}}} \ . 
\end{equation}

By considering the transformation \cite{Arhrib:2011uy}
\begin{equation}\label{eq:tanalpha}
    \tan{2\alpha}= \frac{2{\mathcal{K}^2_3}}{\mathcal{K}_3^2-\mathcal{K}_2^2},
\end{equation}
the physical mass of the heavy Higgs can be rewritten as,

\begin{equation}\label{eq:HeavyMass}
    M_H^2 = \frac{1}{2} \Bigl\{ \mathcal{K}_1^2+\mathcal{K}_2^2 + \sqrt{(\mathcal{K}^2_1-\mathcal{K}_2^2)^2+4\mathcal{K}^4_3} \Bigr\}.
\end{equation}

This implies that the heavy Higgs avoids tachyonic modes when $f(\mu)(1+\frac{\tan^2{\beta}}{2})^{-3/2} > 0$ for a given set of values in $\mathcal{P}_1$, where $f(\mu)$ is a quadratic function of the form $-a\mu^2+b\mu+c$ (see the Appendix \ref{sec:AppB}). Moreover, $f(\mu)>0$ for $\mu \in [\mu_-,\mu_+]$ and the full expression for $\mu_{\pm}$ are given in Eq.~(\ref{eq:widemu}). The minimum limits provided by Eqs.~(\ref{mu1}),~(\ref{mu2}),~(\ref{mu3}) and $\mu_-$ might be mutually overpowering depending on the specific numerical values assigned to $\mathcal{P}_1$, and this aspect should be considered when establishing the lower bound of lepton number violating parameter $\mu$. We select the maximum value among the values yielded by Eqs.~(\ref{mu1}),~(\ref{mu2}),~(\ref{mu3}) and $\mu_-$ thus readjusting the constraints to the form $\mu \in [\mu_L,\mu_+]$ where $\mu_L=\max\{\mu_{1},\mu_{2},\mu_{3},\mu_{-}\}$.

\begin{widetext}
\begin{equation}\label{eq:widemu}
 \mu_{\pm}
 = \frac{M_W S_W}{8\tan{\beta}\sqrt{\pi\alpha_e}(1+\frac{\tan^2{\beta}}{2})}\Bigl[ \lambda + 2(\lambda_1+\lambda_4)\tan^2{\beta}\pm \sqrt{2\lambda^2+8\lambda\tan^2{\beta}(\lambda_1+\lambda_4+\lambda_2\tan^2{\beta}+\lambda_3\tan^2{\beta})} \Bigr]
\end{equation}
\end{widetext}

This provides an upper bound for $\mu$ when defining the parameter space and ensures the absence of tachyonic modes for the heavy Higgs in the HTM.

An upper limit for $\tan{\beta}$ can be derived by examining electroweak precision measurements. In the Standard Model, the presence of custodial symmetry ensures that $\rho=1$ at the tree level, whereas in the Higgs triplet model, the relation becomes

\begin{equation}\label{eq:rho}
    \rho = \frac{1+\frac{2v^2_{\Delta}}{v_{\Phi}^2}}{1+\frac{4v^2_{\Delta}}{v_{\Phi}^2}} = \frac{1+\frac{1}{2}\tan^2{\beta}}{1+\tan^2{\beta}}.
\end{equation}

The experimental value of the rho parameter, $\rho^{exp}=1.0008^{+0.0017}_{-0.0007}$,\cite{Aoki:2012yt}, being close to unity, leads to the bound $\tan{\beta} \lesssim 0.0633$.

\section{The Details of the Calculation}
\label{sec:method}

\begin{figure*}
\includegraphics[width=0.8\linewidth]{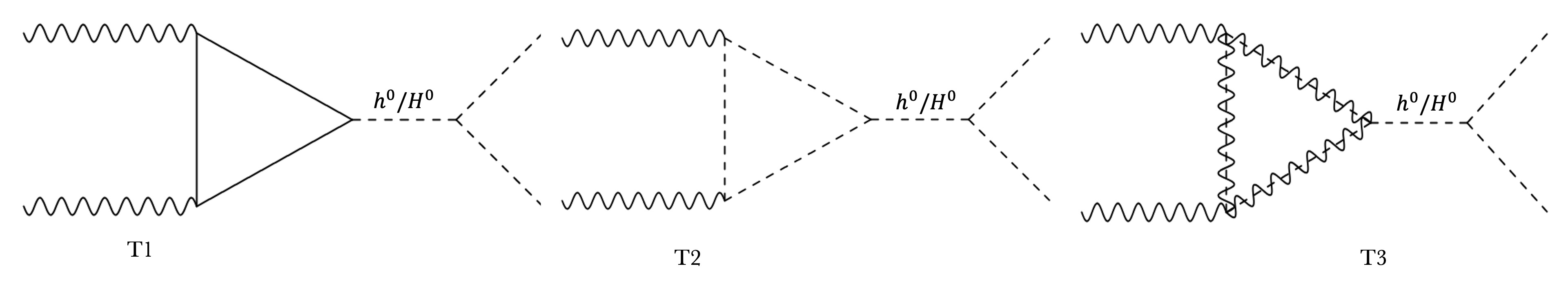}% Here is how to import EPS art
\caption{\label{Feyn1} Triangle-type diagrams contribute to the one-loop level process $\gamma\gamma \rightarrow \phi\phi$, where $\phi$ can be either $h^0$ or $A^0$. In these diagrams, solid lines denote Standard Model fermions, specifically in the T1 topology. Within the loops, dashed lines represent charged Higgs bosons ($H^{\pm},H^{\pm\pm}$), while the wavy lines in the loops represent W bosons. }
\end{figure*}

\begin{figure*}
\includegraphics[width=0.8\linewidth]{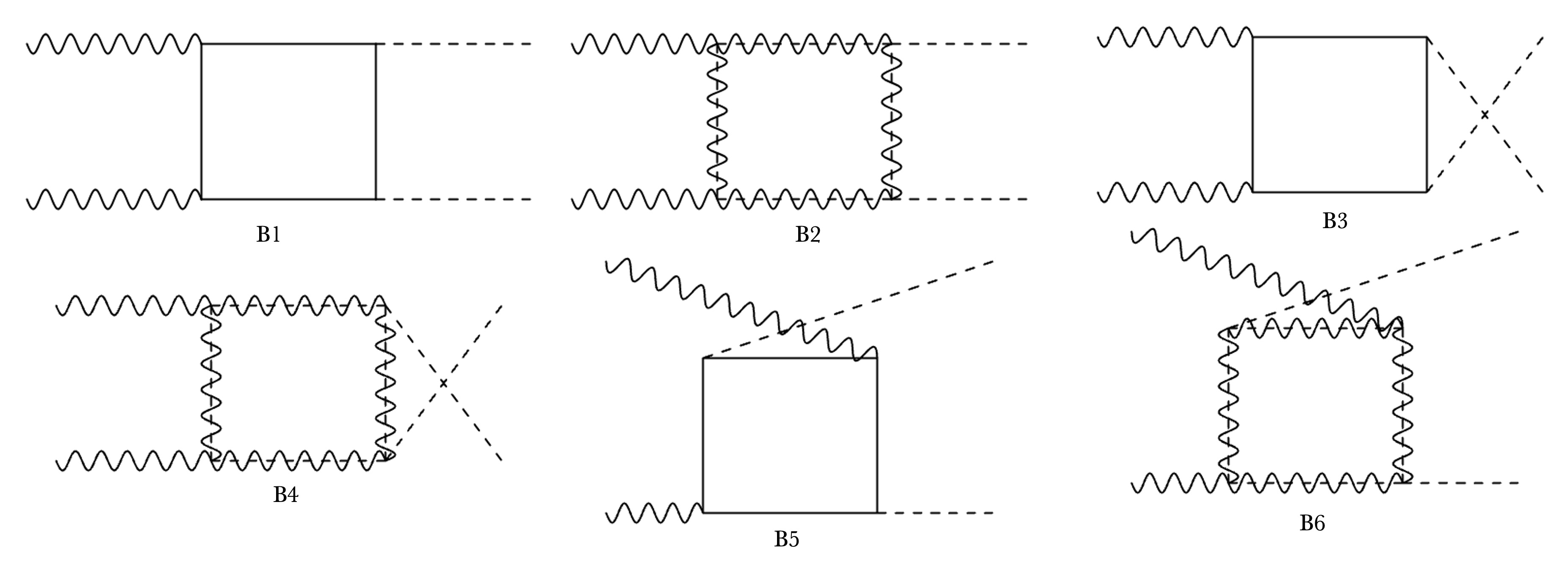}% Here is how to import EPS art
\caption{\label{Feyn2}Box-type diagrams contribute to the one-loop level process $\gamma\gamma \rightarrow \phi\phi$, where $\phi$ can be either $h^0$ or $A^0$. In these diagrams, solid lines denote Standard Model fermions, specifically in the B1, B3 and B5 topologies. Within the loops, dashed lines represent charged Higgs bosons ($H^{\pm},H^{\pm\pm}$), and the charged Goldstone boson $G^{+}$, while the wavy lines in the loops represent W bosons. }
\end{figure*}

\begin{figure*}
\includegraphics[width=0.8\linewidth]{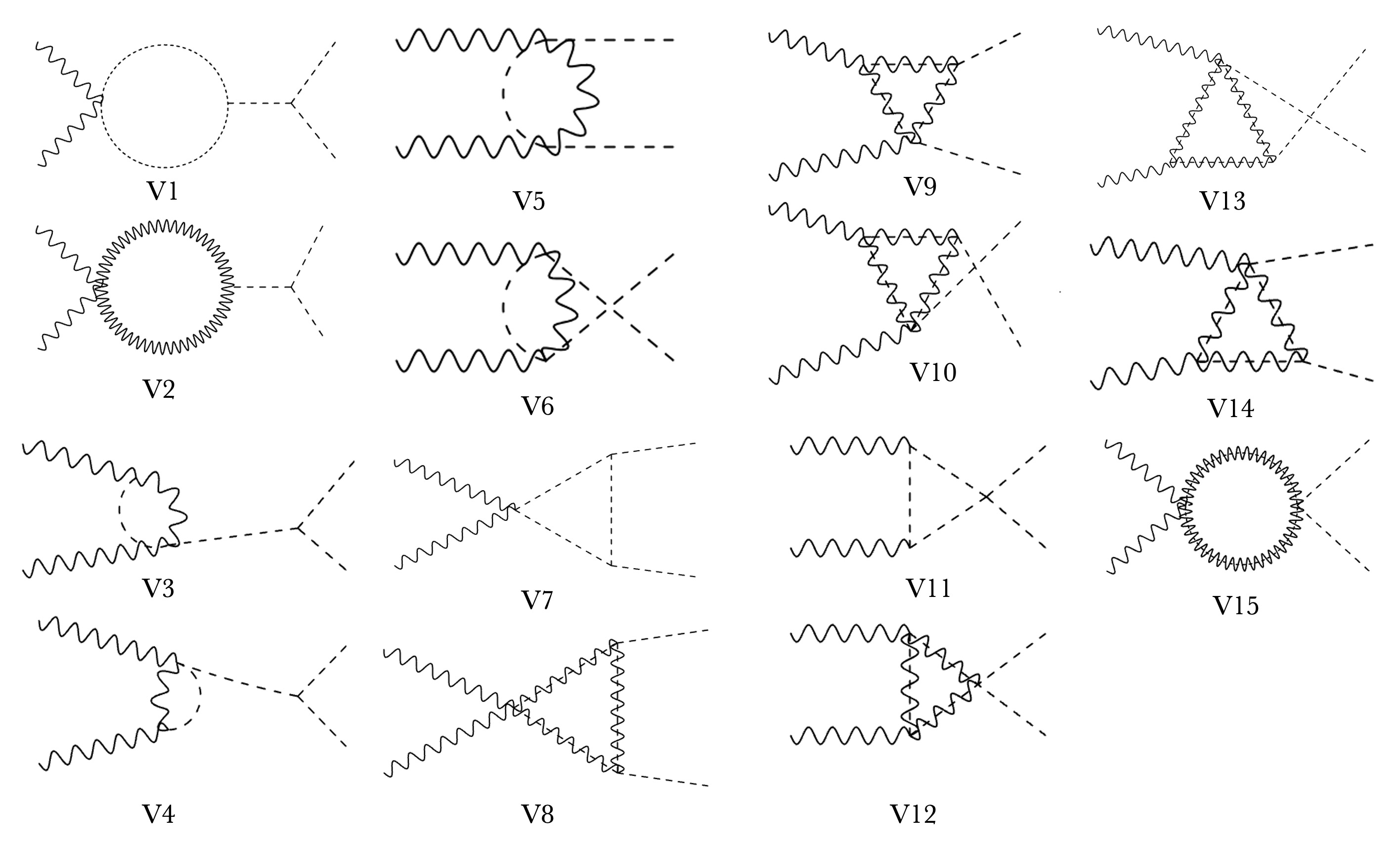}
\caption{\label{Feyn3}Quartic-type diagrams contribute to the one-loop level process $\gamma\gamma \rightarrow \phi\phi$, where $\phi$ can be either $h^0$ or $A^0$. In these diagrams,dashed lines Within the loops  represent charged Higgs bosons ($H^{\pm},H^{\pm\pm}$), and the charged Goldstone boson $G^{+}$, while the wavy lines in the loops represent W bosons. }
\end{figure*}

The process of the Higgs pair production in photon collision is denoted by

\begin{equation}\label{eq:AmuAnu}
    A_{\mu}(k_1)+A_{\nu}(k_2) \rightarrow \phi(k_3)+\phi(k_4),
\end{equation}
where $\phi \in \{h,A^0\}$ and their corresponding 4-momenta are enclosed in parentheses. The one-loop Feynman diagrams for $\gamma\gamma \rightarrow \phi\phi$ can be categorized into triangle-type (Fig.~\ref{Feyn1}), box-type (Fig.~\ref{Feyn2}), and quartic coupling-type (Fig.~\ref{Feyn3}) diagrams. Since the tensor amplitude for the process $\gamma\gamma \rightarrow \phi\phi$ at the one-loop level is computed by summing all unrenormalized reducible and irreducible contributions, the resulting values are finite and maintain gauge invariance. The tensor amplitude and the  amplitude are expressed as follows:
\begin{equation}\label{eq:tensoraAmp}
   \mathcal{M}_{\mu\nu} =(\mathcal{M}_{\mu\nu}^{box}+\mathcal{M}_{\mu\nu}^{triangle}+\mathcal{M}_{\mu\nu}^{quartic}) \ ,
\end{equation}
\begin{equation}\label{eq:invarAmp}
    \mathcal{M}= \mathcal{M}_{\mu\nu}\epsilon^{\mu}(k_1,\lambda_1)\epsilon^{\nu}(k_2,\lambda_2) \ .
\end{equation}

The total partonic cross sections for $\gamma\gamma \rightarrow \phi\phi$ processes are expressed as

\begin{equation}\label{eq:partXsections}
\hat{\sigma}(\hat{s},\gamma\gamma \rightarrow \phi\phi)=\frac{1}{32\pi \hat{s}^2}\int_{\hat{t^-}}^{\hat{t^+}}dt\sum_{spins}|\mathcal{M}|^2,
\end{equation}

where $\hat{t^{\pm}}=(M_{\phi}^2-\hat{s}/2)\pm \sqrt{(\hat{s}/2-M_{\phi}^2)^2-M_{\phi}^4}$. Since the Higgs pair production via photon-photon collisions is a subprocess of $\mu^+\mu^-$ collisions at the muon collider, the total cross section of this process can be conveniently obtained by utilizing the expression

\begin{equation}\label{fullXsec}
\sigma(s,\mu^+\mu^- \rightarrow \gamma \gamma \rightarrow \phi \phi) = \int^{1}_{\frac{2M_{\phi}^2}{\sqrt{s}}}d\tau \frac{d\mathcal{L}_{\gamma\gamma}}{d\tau}\hat{\sigma}(\hat{s}=\tau s,\gamma\gamma \rightarrow \phi\phi),
\end{equation}
along with the photon luminosity
\begin{equation}\label{LuminosityFunc}
\frac{d\mathcal{L}_{\gamma\gamma}}{d\tau}=  \int^{1}_{\tau}\frac{dx}{x}f_{\gamma/ \mu}(x)f_{\gamma/ \mu}(\frac{\tau}{x}),
\end{equation}
where $\sqrt{\hat{s}}$ and $\sqrt{s}$ represent the center-of-mass energies of $\gamma\gamma$ and $\mu^+\mu^-$ collisions, respectively.

To calculate the total cross sections, we employed the Weizsäcker-Williams approximation, which represents the leading order (LO) contribution for the photon PDFs, as shown in Eq.~(\ref{eq:EPALO}). 
In the work presented in Ref.~\cite{Garosi:2023bvq}, the solution to the DGLAP equations has been achieved through iterative techniques. As a second approach in performing the cross section calculations, we use their second order corrected PDF, denoted as LO$ + \mathcal{O}(\alpha_e^2t^2)$ as well, see the Eq.~(\ref{LLf}).
Third, we used EW PDF at the leading -log (LL) accuracy available at \cite{Han:2020uid}.
 
In this paper, the explicit presentation of the matrix element expressions has been omitted due to their length. We implemented the HTM Lagrangian into {\tt FeynRules}\footnote{Even though there is a FeynRules model file available at Ref.~\cite{Fuks:2019clu}, we have implemented our own FeynRules model file.}. The generation of the {\tt FeynArts} models files was performed by {\tt FeynRules} \cite{Christensen:2008py, Alloul:2013bka}. The generation of one-loop amplitudes was performed by {\tt FeynArts} \cite{Hahn:2000kx} and the subsequent generation of the matrix element squared was performed using {\tt FormCalc} \cite{Hahn:1998yk, Denner:1991kt}. In order to incorporate photon structure functions, distinct Fortran subroutines were developed for both LO and the second order corrected EPA. The numerical assessments of the integration over the $2 \rightarrow 2$ phase space were carried out using the {\tt CUBA} library\cite{Hahn:2016ktb,Hahn:2004fe}.

 
\section{Analysis, Results and Discussion}
\label{sec:results}

Initially, we examine the overall rates of the partonic processes $\gamma\gamma \rightarrow h^0h^0, A^0A^0$ in the center of mass system of $\gamma\gamma$ before convoluting them with the photon energy spectrum in a muon collider. In our numerical analysis, we employed $M_W = 80.379 \ \mathrm{GeV}$, $M_Z = 91.18 \ \mathrm{GeV}$, the Weinberg angle $S_W = \sqrt{1 - \frac{M_W^2}{M_Z^2}}$, and the fine structure constant $\alpha_e = 1/137.03598$.

According to BFB conditions, Ref.~\cite{Arhrib:2011uy}, the parameters $\lambda_i$, ($i=1,2,3,4$) can be written as functions of masses of CP-odd and CP-even Bosons. 
The requirement for the square root expressions in Eq.~(\ref{eq:b2}) and Eq.~(\ref{eq:b3}) to be real is consistent with the conditions that $\lambda_2 + \lambda_3$ and $\lambda_2 + \lambda_3/2$ are both positive. We can derive the following expressions to identify the range of parameter values that satisfy the condition of vacuum stability: 
\begin{equation}\label{eq:MAless}
    M_{A^0} \leq \frac{(v^2_{\Phi}+4v_{\Delta}^2)^{1/2}}{v_{\Phi}}(\cos{\alpha}^2M_{H^0}^2+\sin{\alpha}^2M_{h^0}^2)^{1/2} \ , 
\end{equation}
\begin{equation}\label{MHpless}
   M_{H^\pm} \leq \frac{(v^2_{\Phi}+2v_{\Delta}^2)^{1/2}}{\sqrt{2}v_{\Phi}}(M^2_{H^{\pm\pm}}+\cos{\alpha}^2M_{H^0}^2+\sin{\alpha}^2M_{h^0}^2)^{1/2} \ . 
 \end{equation}
 
Throughout the calculations, we explore various hierarchies between $M_{H^{\pm}}$ and $M_{H^{\pm\pm}}$, primarily determined by the sign of $\lambda_4$ at low $\tan{\beta}$ values, as explained by Eq.~(\ref{eq:massDiff}) given below 
\begin{equation}\label{eq:massDiff}
    M_{H^{\pm}}^2-M_{H^{\pm\pm}}^2=\frac{M^2_W S^2_W}{4\alpha_e\pi}\lambda_4+\frac{\mu M_W S_W}{\sqrt{2\pi\alpha_e}}\tan{\beta}+\mathcal{O}({\tan^2{\beta}}) \ .
\end{equation}
The mass difference between $H^\pm$ and $H^{\pm\pm}$, denoted as $\Delta M = M_{H^\pm} - M_{H^{\pm\pm}}$, is proportional to the coupling constant $\lambda_4$. When $\lambda_4$ is positive at small $v_{\Delta}$ or $\tan{\beta}$, the mass of the singly charged Higgs boson tends to be larger than that of the doubly charged Higgs boson. On the other hand, if $\lambda_4$ is negative it can lead to the opposite scenario where the mass of $H^{\pm}$ is smaller than that of $H^{\pm\pm}$.

\begin{figure}[b]
    \centering
    \includegraphics[width=1.01\linewidth]{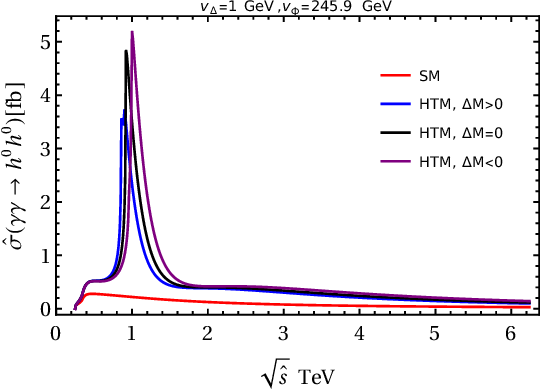} 
    \caption{The partonic cross section of $\gamma\gamma \rightarrow h^0h^0$ as a function of $\sqrt{\hat{s}}$ for different values of $M_h$,$M_H$,$M_{A^0}$, $M_{H^\pm}$ and $M_{H^{\pm\pm}}$. }
    \label{part1}
\end{figure}

\begin{figure}[b]
    \centering
    \includegraphics[width=1.01\linewidth]{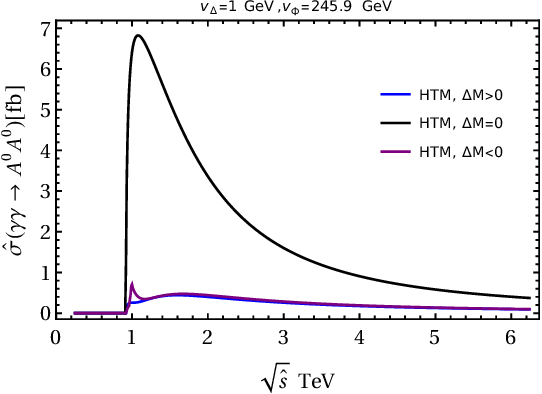} 
    \caption{ The partonic cross section of $\gamma\gamma \rightarrow A^0A^0$ as a function of $\sqrt{\hat{s}}$ for different values of $M_h$,$M_H$,$M_{A^0}$, $M_{H^\pm}$ and $M_{H^{\pm\pm}}$. }\label{part2}
\end{figure}

At $v_{\Delta}/v_{\Phi}\ll1$ and $\alpha \approx 0$, the Eq.~(\ref{eq:MAless}) is reduced to
 $M_{A^0}\lessapprox M_{H^0}$. In our analysis, we considered three distinct scenarios based on the sign of $\Delta M$:
\begin{enumerate}
    \item Utilizing input values of $M_{H^0} = 463.5$ GeV, $M_{H^{\pm\pm}}= 428.31$ GeV, $v_{\Phi}=245.9$ GeV, $v_{\Delta}=1$ GeV, and $\cos{\alpha}=0.999$, we determined the maximum value of $M_{A^0}$ to be approximately 463.5 GeV, with a corresponding maximum value of $M_{H^{\pm}}$ at 446.28 GeV. This parameter set was employed to compute partonic cross sections for the scenario where $\Delta M > 0$.
   \item For the case where $\Delta M < 0$, we employed input values of $M_{H^0} = 463.5$ GeV, $M_{H^{\pm\pm}}= 496.26$ GeV, $v_{\Phi}=245.9$ GeV, $v_{\Delta}=1$ GeV, and $\cos{\alpha}=0.999$. The resulting analysis determined the maximum values of $M_{A^0}$ and $M_{H^{\pm}}$ to be approximately 463.5 GeV and 480.19 GeV, respectively.
  \item In the scenario where $\Delta M= 0$, input values of $M_{H^0} = 459.14$ GeV, $M_{H^{\pm\pm}}= 459.14$ GeV, $v_{\Phi}=245.9$ GeV, $v_{\Delta}=1$ GeV, and $\cos{\alpha}=0.999$ were employed. The maximum possible values for $M_{A^0}$ and $M_{H^{\pm}}$ were found to be approximately 459.15 GeV. 
\end{enumerate}
Subsequently, we employed this parameter set to compute partonic cross sections for $\gamma\gamma \rightarrow h^0h^0$ within the energy range $2M_h \leq \sqrt{\hat{s}} \leq 6$ TeV, as illustrated in Fig.~\ref{part1}. Comparing the cross section profiles between the Higgs triplet model and Standard Model, the HTM reaches peak values of 3.8 fb for $\Delta M>0$, 4.8 fb for $\Delta M=0$, and 5.2 fb for $\Delta M<0$.
Utilizing the same parameter sets, we calculated the partonic cross sections for $\gamma\gamma \rightarrow A^0A^0$, and the results are presented in Fig.~\ref{part2}. The cross sections attain peak values of 0.6 fb for $\Delta M>0$, 6.8 fb for $\Delta M=0$, and 0.8 fb for $\Delta M<0$.

\subsection{$\mu^+\mu^-\rightarrow \gamma\gamma \rightarrow h^{0}h^{0},A^{0}A^{0}$ cross sections}
\label{subsec:xsecs}

During the process of performing computations, utilizing the mass parameters within the parameter space $\mathcal{P}_2$ may not be the most efficient approach, as it could potentially lead to violations of unitarity conditions \cite{Arhrib:2011uy}. As a result, in order to ensure the validity and consistency of the computed convoluted cross sections, we proceed by employing the inputs that reside within the parameter space $\mathcal{P}_1$. This strategic choice helps to maintain the integrity of the calculations and supports accurate predictions in accordance with the theoretical framework.

\begin{figure}
    %\centering
    \includegraphics[width=1.01\linewidth]{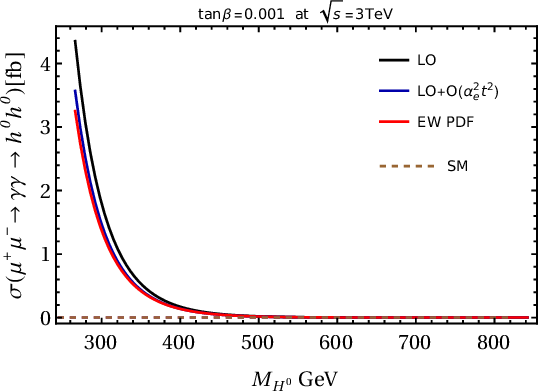} % Replace "figure.png" with the actual file name and extension
    \caption{The total cross section $\sigma(\mu^+\mu^- \rightarrow \gamma\gamma \rightarrow h^0h^0)$ as a function of $M_{H^0}$  values for $M_{H^{\pm\pm}} \approx M_{H^\pm}$ . }
    \label{x1}
\end{figure}

\begin{figure}
    %\centering
    \includegraphics[width=1.01\linewidth]{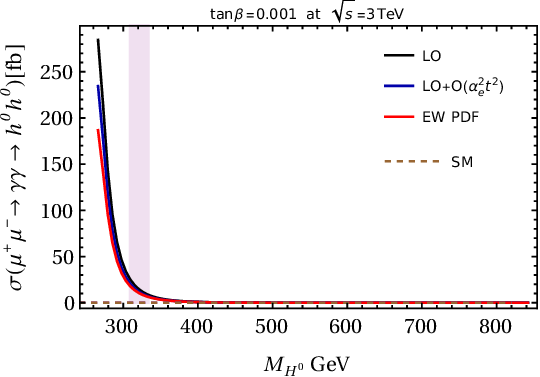} % Replace "figure.png" with the actual file name and extension
    \caption{ The total cross section $\sigma(\mu^+\mu^- \rightarrow \gamma\gamma \rightarrow h^0h^0)$ as a function of $M_{H^0}$  values for $M_{H^{\pm\pm}} < M_{H^\pm}$. The pink region in the plot represents the scenario where $M_{H^{\pm\pm}}$ falls within exclusion limits between 200 GeV and 220 GeV, resulting in corresponding $M_{H^0}$ mass values ranging from 315 GeV to 335 GeV.}\label{x2}
\end{figure}

\begin{figure}
    %\centering
    \includegraphics[width=1.01\linewidth]{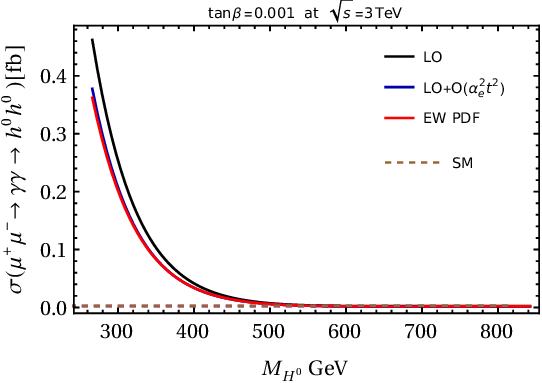} % Replace "figure.png" with the actual file name and extension
    \caption{The total cross section $\sigma(\mu^+\mu^- \rightarrow \gamma\gamma \rightarrow h^0h^0)$ as a function of $M_{H^0}$  values for $M_{H^{\pm\pm}} > M_{H^\pm}$ .}\label{x3}
\end{figure}

\begin{figure}
    %\centering
    \includegraphics[width=1.01\linewidth]{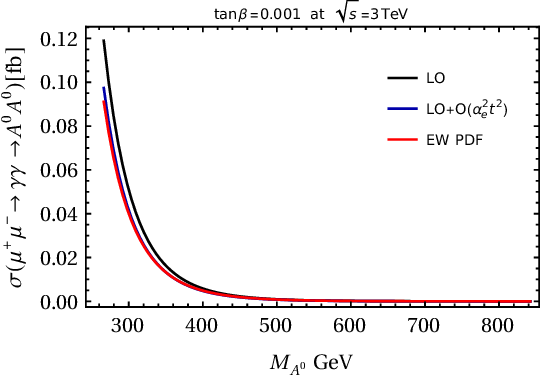} % Replace "figure.png" with the actual file name and extension
    \caption{The total cross section $\sigma(\mu^+\mu^- \rightarrow \gamma\gamma \rightarrow A^0A^0)$ as a function of $M_{A^0}$  values for $M_{H^{\pm\pm}} \approx M_{H^\pm}$ . }\label{x4}
\end{figure}

\begin{figure}
    %\centering
    \includegraphics[width=1.01\linewidth]{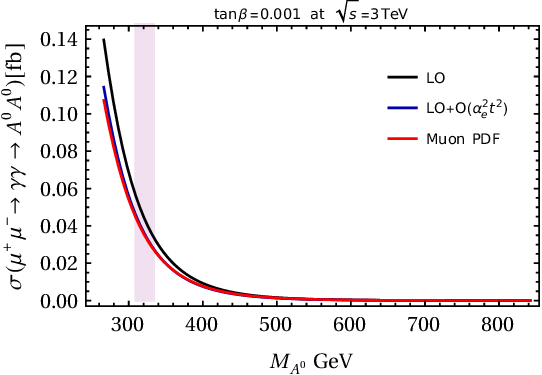} % Replace "figure.png" with the actual file name and extension
    \caption{The total cross section $\sigma(\mu^+\mu^- \rightarrow \gamma\gamma \rightarrow A^0A^0)$ as a function of $M_{A^0}$  values for $M_{H^{\pm\pm}} < M_{H^\pm}$ .  The pink region in the plot represents the scenario where $M_{H^{\pm\pm}}$ falls within exclusion limits between 200 GeV and 220 GeV, resulting in corresponding $M_{A^0}$ mass values ranging from 315 GeV to 335 GeV. }\label{x5}
\end{figure}

\begin{figure}
    %\centering
    \includegraphics[width=1.01\linewidth]{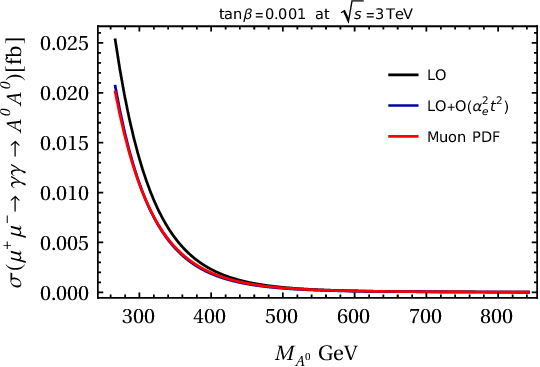} % Replace "figure.png" with the actual file name and extension
    \caption{ The total cross section $\sigma(\mu^+\mu^- \rightarrow \gamma\gamma \rightarrow A^0A^0)$ as a function of $M_{A^0}$  values for $M_{H^{\pm\pm}} > M_{H^\pm}$. }\label{x6}
\end{figure}

In Fig.~\ref{x1}- Fig.~\ref{x6}, at $\sqrt{s}=3$ TeV, we present the cross sections for various input values of $\lambda_4$, while maintaining fixed values of $\lambda=0.51$, $\lambda_1=10$, $\lambda_2=1$, $\lambda_3=-1$, $\alpha \approx 0$, and $\tan{\beta}=0.001$. With $\lambda_4$ under variation, the associated upper and lower bounds for the lepton number violating parameter undergo changes in accordance with Eqs.~( \ref{mu1}),~(\ref{mu2}),~(\ref{mu3}) and (\ref{eq:widemu}) and considerations related to the avoidance of tachyonic modes. Nevertheless, for computational efficiency, we set an arbitrary upper limit of 2 for $\mu$, although it can extend to much larger values ( $\approx 5 \times 10^3$GeV) given by Eq.~(\ref{eq:widemu}). Additionally, Table \ref{tab1} provides the generated values of $M_{A^0}$ and $M_{H^0}$ corresponding to these input parameters.

\vspace{0.01mm}


\begin{figure}
    %\centering
    \includegraphics[width=1.01\linewidth]{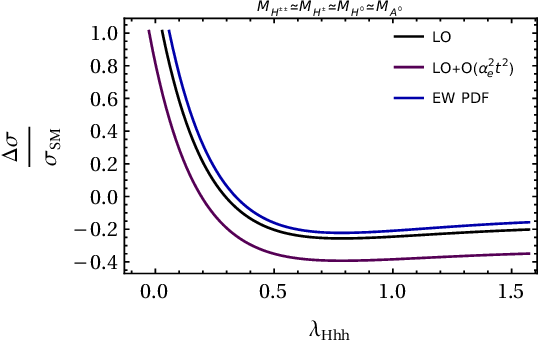} % Replace "figure.png" with the actual file name and extension
    \caption{The relative difference between the cross sections of Higgs boson pair production in the HTM and the SM, as a function of the trilinear scalar coupling $H^0h^0h^0$, is analyzed at a center-of-mass energy of $\sqrt{s}=3 \ \mathrm{TeV}$. }\label{ratio1}
\end{figure}

\begin{figure}
    %\centering
    \includegraphics[width=1.01\linewidth]{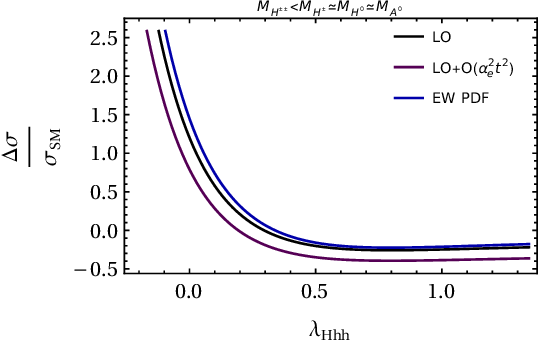} % Replace "figure.png" with the actual file name and extension
    \caption{The relative difference between the cross sections of Higgs boson pair production in the HTM and the SM, as a function of the trilinear scalar coupling $H^0h^0h^0$, is analyzed at a center-of-mass energy of $\sqrt{s}=3 \ \mathrm{TeV}$. }\label{ratio2}
\end{figure}

\begin{figure}
    %\centering
    \includegraphics[width=1.01\linewidth]{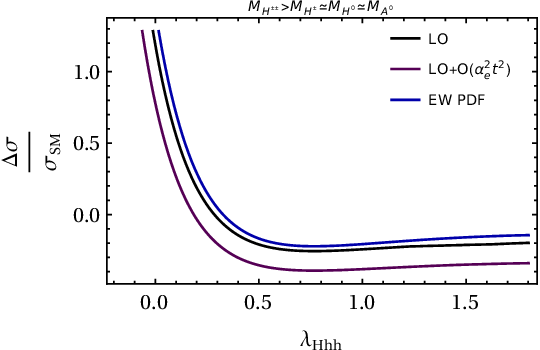} % Replace "figure.png" with the actual file name and extension
    \caption{The relative difference between the cross sections of Higgs boson pair production in the HTM and the SM, as a function of the trilinear scalar coupling $H^0h^0h^0$, is analyzed at a center-of-mass energy of $\sqrt{s}=3 \ \mathrm{TeV}$.  }\label{ratio3}
\end{figure}

\begin{table}[b]
\caption{\label{tab1}
The presented mass ranges in this table correspond to the input values $\tan{\beta}=0.001$, $\lambda=0.51$, $\lambda_1=10$, $\lambda_2=1$, $\lambda_3=-1$, and $\mu=[0.2,2]$. The numerical values demonstrate that $\Delta M > 0$ for $\lambda_4 > 0$, $\Delta M = 0$ for $\lambda_4 = 0$, and $\Delta M < 0$ for $\lambda_4 < 0$.}
\begin{ruledtabular}
\begin{tabular}{ccccc}
     & $M_{H^{\pm}}\mathrm{GeV}$ & $M_{H^{\pm\pm}}\mathrm{GeV}$ & $M_{H^0}\mathrm{GeV}$ & $M_{A^0}\mathrm{GeV}$ \\
    \hline
    $\lambda_4=+1.82$ & 201-809.6 & 114-792.5 & 261-826.5& 261-826.5\\
    
    $\lambda_4=0$& 261-827& 261-827 & 261-827 & 261-827 \\
    
    $\lambda_4=-1.82$&309-843& 351-859 & 261-826.5& 261-826.5 \\
\end{tabular}
\end{ruledtabular}
\end{table}

For $\tan{\beta} \geq 8.13\times 10^{-7}$, the branching ratio $Br(H^{\pm\pm} \rightarrow W^{\pm}W^{\pm}) \approx 1$ is greater than $Br(H^{\pm\pm} \rightarrow l^{\pm}l^{\pm})$. The ATLAS experiment has excluded $H^{\pm\pm}$ boson masses between 200 GeV and 220 GeV at a 95\% confidence level, Ref.~\cite{Rahili:2019ixf}. We encounter these exclusion limits at 
$\lambda_4 = +1.82$ , and the corresponding masses for $M_{A^0}$ and $M_{H^0}$ are considered when calculating the cross sections in this scenario.

In computing the cross sections of $\mu^+\mu^- \rightarrow \gamma \gamma \rightarrow \phi\phi$,
our first approach involved convoluting the partonic cross sections with the Weizsäcker-Williams approximation (LO), followed by the second approach, where we convoluted with the Weizsäcker-Williams approximation corrected up to the second order, $\mathcal{O}(\alpha_e^2t^2)$. Lastly, we employed the EW PDF sets, and the results were systematically compared. The percentage differences and comparisons in cross sections for $h^0h^0$ and $A^0A^0$ production processes using the three approaches can be summarized as follows.
\begin{enumerate}
    \item According to Figs.~\ref{x1}, \ref{x2}, and \ref{x3}, the cross section for $\mu^+\mu^- \rightarrow \gamma \gamma \rightarrow h^0h^0$ at $M_{H^{\pm}}\approx M_{H^{\pm\pm}}$ is approximately $10$ times larger than the cross section when $M_{H^{\pm\pm}}>M_{H^{\pm}}$, and it is approximately $60$ times larger when $M_{H^{\pm}}> M_{H^{\pm\pm}}$ compared to $M_{H^{\pm\pm}}\approx M_{H^{\pm}}$.
    \item The cross section for $\mu^+\mu^- \rightarrow \gamma \gamma \rightarrow h^0h^0$ obtained from the LO approximation exhibits an approximate 6.5\% difference compared to the results obtained using EW PDF. Furthermore, the difference between the results obtained from LO+$\mathcal{O}(\alpha_e^2t^2)$ and EW PDF is approximately 1.12\%. The difference between the results obtained from LO and LO+$\mathcal{O}(\alpha_e^2t^2)$ stands at around 5\%.
    \item In Figs.~\ref{x4},~\ref{x5}, and \ref{x6}, the cross section for $\mu^+\mu^- \rightarrow \gamma \gamma \rightarrow A^0A^0$ is notably higher, being approximately $1.2$ times larger at $M_{H^{\pm}}> M_{H^{\pm\pm}}$ compared to the scenario where $M_{H^{\pm}}\approx M_{H^{\pm\pm}}$. Similarly, when $M_{H^{\pm}}>M_{H^{\pm\pm}}$, the cross section is approximately $5.6$ times larger than the case when $M_{H^{\pm}}<M_{H^{\pm\pm}}$.
    \item In the mass range of $300 \ \mathrm{GeV} \leq M_{A^0} \leq 450 \
    \mathrm{GeV}$, the cross section for $\mu^+\mu^- \rightarrow \gamma \gamma \rightarrow A^0A^0$ shows a 6\% discrepancy between the leading order approximation results and results obtained using EW PDF. The contrast between LO+$\mathcal{O}(\alpha_e^2t^2)$ and EW PDF is approximately 1\%, with the difference between LO and LO+$\mathcal{O}(\alpha_e^2t^2)$ results remaining at around 5\%. 
\end{enumerate}

\subsection{Decoupling limits}
\label{subsec:limits}

In the decoupling limit, characterized by $\alpha \rightarrow 0$, the lighter scalar, often identified as the SM-like Higgs boson ($h^0$), retains properties similar to the SM Higgs boson. Meanwhile, the heavier scalar masses become decoupled from the electroweak scale. As $\alpha$ approaches zero, the tree level scalar trilinear couplings can be expressed as series of $\tan{\beta}$, Eqs.~(\ref{cup1})- (\ref{cup10}), providing insight into the decoupling behavior at specific values. By defining, 

 \begin{equation}\nonumber\label{eq:lamPrime}
        \lambda^{'} = -\frac{3\sqrt{\pi\alpha_{e}}}{vM_WS_W}\Big(\frac{M_h^2}{-\frac{3}{2}+\frac{3}{8}\tan^2{\beta}+\mathcal{O}(\tan^4{\beta})}\Big),
    \end{equation}
  we observe that $ \lambda^{HTM}_{hhh}\Big|_{\lambda = \lambda^{'}}\rightarrow \lambda_{hhh}^{SM}$. According to our numerical inputs, $\lambda^{HTM}_{h^0h^0h^0}$ and $\lambda^{SM}_{h^0h^0h^0}$ are nearly indistinguishable, with the percentage difference being approximately 1\%. This relationship is governed by the expression shown in Eq.~(\ref{cup1}).

  The $H^0h^0h^0$ trilinear coupling at $\alpha \rightarrow 0$, see the Eq.~(\ref{cup2}), is meeting a weak-coupling decoupling \cite{Haber:2013mia} at

\begin{equation}
\mu^{d} \approx \frac{M_WS_W(\lambda_1+\lambda_4)}{\sqrt{8\pi\alpha_{e}}}\tan{\beta}.
\end{equation}

Here, $\mu^d$ represents the associated lepton flavor violating parameter value at which $\lambda^{HTM}_{Hhh}$ converges to zero  ($\lambda^{HTM}_{Hhh} \Big|_{\mu=\mu^{d}}\rightarrow 0$) for $\tan{\beta}\ll1$. The Table \ref{tab2} displays the weak-coupling decoupling scenario, along with $M_{H^0}$, for each hierarchy case considered in the aforementioned calculations, where the coupling $H^0h^0h^0$ is found to approach small values. 

\begin{table}
\caption{\label{tab2}The table illustrates the mass values of $M_{H^0}$ and the trilinear Higgs coupling corresponding to different $\lambda_4$ inputs in the weak-coupling decoupling scenario.}
\begin{ruledtabular}
\begin{tabular}{cccc}
    $\lambda_4$ & $\mu^{d}$  & $M_{H^0}(\mathrm{GeV})$ & $\lambda_{Hhh}^{HTM}$  \\
  \hline $+1.82$  & $1.047$ & $597.98$ &  $10^{-6}$ \\
   $0$      & $0.869$ & $544.78$ & $10^{-7}$ \\
   $-1.82$  & $0.725$ & $497.60$ & $10^{-7}$
\end{tabular}
\end{ruledtabular}
\end{table}

The couplings $h^0H^+H^-$ and $h^0h^0H^{+}H^{-}$ demonstrate sensitivity to distinct sets of Feynman diagrams. Specifically, $h^0H^+H^-$ exhibits sensitivity to T2, B2, B4, B6, V1, and V7, while $h^0h^0H^{+}H^{-}$ is responsive to the topologies V11 and V15 depicted in Figs.~\ref{Feyn1}, \ref{Feyn2}, and \ref{Feyn3}. Examining Eq.~(\ref{cup4}) and Eq.~(\ref{cup6}), it is observed that the magnitudes of these couplings reach their maximum values when the parameter $\lambda_4$ is set to $1.82$, while their minimum values are obtained at $\lambda_4=-1.82$ in the carried out computations. Notably, the results presented in Figs.~\ref{x1},~\ref{x2}, and ~\ref{x3} suggest that the cross sections of $h^0h^0$ production experience enhancement when $\lambda_4 > 0$, despite the small value of $\lambda_{H^0h^0h^0}^{HTM}$ at the weak coupling decoupling limit. Additionally, the coupling $h^0h^0H^{++}H^{--}$ shows sensitivity to the Feynman diagrams V11 and V15 and Eq.~(\ref{cup7}) indicates this coupling maintains a constant and nonzero value throughout the calculations.

The couplings $h^0A^0A^0$ and $H^0A^0A^0$ exhibit sensitivity to specific Feynman diagrams, namely T1, T2, T3, V1, and V2, as illustrated in Fig.~\ref{Feyn1} and Fig.~\ref{Feyn3}. According to Eq.~(\ref{cup3}), as $\lambda^{HTM}_{H^0A^0A^0} \approx \sqrt{2}\mu\tan^2{\beta}$ for small $\mu$, $\lambda^{HTM}_{H^0A^0A^0}$ also approaches the weak-coupling decoupling limit when $\tan{\beta} \ll 1$ and $\lambda_2 + \lambda_3 = 1 + (-1) = 0$. For the values of $\mu\in [0.2,2]$, $2.8\times 10^{-5}<\lambda^{HTM}_{H^0A^0A^0}<2.8\times10^{-3}$, but the $H^0A^0A^0$ coupling becomes substantial for large $\mu$. Despite the relatively small magnitude of $H^0A^0A^0$ in the calculations, Eq.~(\ref{cup8}) reveals that the magnitude of $h^0A^0A^0$ reaches its maximum when $\lambda_4$ is greater than 0. Conversely, this coupling attains its minimum values at $\lambda_4=-1.82$ in our calculations. Moving on to the couplings $A^0A^0H^{++}H^{--}$ and $A^0A^0H^{+}H^{-}$, their sensitivity lies in Feynman diagrams labeled as V11 and V15 in Fig.~\ref{Feyn3}. According to Eq.~(\ref{cup9}), $\lambda_{A^0A^0H^{++}H^{--}}^{HTM}$ remains nonzero, while Eq.~(\ref{cup10}) indicates that $\lambda_{A^0A^0H^{+}H^{-}}^{HTM} \approx -3/2(\lambda_1+\lambda_4)\tan^2{\beta}$ is relatively small. Notably, the results presented in Figs.~\ref{x4},~\ref{x5}, and ~\ref{x6} suggest an enhancement in the cross sections of $A^0A^0$ production when $\lambda_4$ is greater than 0. 

In comparing the cross sections of $h^0h^0$ production between the HTM and the SM, we adopt a relative difference approach expressed as follows:
\begin{align}\label{rel1}
\frac{\Delta\sigma}{\sigma_{SM}}= \frac{\sigma_{HTM}(\mu^+\mu^-\rightarrow\gamma\gamma \rightarrow h^0h^0)}{\sigma_{SM}(\mu^+\mu^-\rightarrow\gamma\gamma \rightarrow h^0h^0)}-1
\end{align}
This relative difference provides a measure of how the HTM cross sections deviate from their SM results. When the relative difference approaches zero, it indicates that the cross sections within the HTM are close to the SM results. The closeness of these results does not depend on the hierarchy of charge and doubly charged Higgs bosons but rather on the choice of parton distribution functions. Particularly, Figs.~\ref{ratio1},~\ref{ratio2}, and~\ref{ratio3} visually depict that the cross sections within the HTM closely align with the SM results for the same values of $\lambda^{HTM}_{H^0h^0h^0}$, irrespective of the hierarchy, as detailed in Table~\ref{tab3}. 

\begin{table}
\caption{\label{tab3}Comparative analysis of $h^0h^0$ production cross sections in the HTM and SM. The table illustrates the values of $\lambda^{HTM}_{H^0h^0h^0}$ at which the relative difference, as defined by Eq.~(\ref{rel1}), approaches zero. The results remain robust across different hierarchies of charge and doubly charged Higgs bosons.}
\begin{ruledtabular}
\begin{tabular}{c|ccc}
&\multicolumn{2}{c}{$\lambda^{HTM}_{H^0h^0h^0}$}\\ \hline
    \backslashbox{Hierarchy}{PDFs} & LO & LO+$\mathcal{O}(\alpha_e^2t^2)$ & EW PDF  \\
  \hline $M_{H^{\pm}}\approx M_{H^{\pm\pm}}$& 0.3 & 0.2 &  0.35 \\
   $M_{H^{\pm}}>M_{H^{\pm\pm}}$  & 0.3 & 0.2 & 0.35 \\
   $M_{H^{\pm}}<M_{H^{\pm\pm}}$  & 0.3& 0.2 & 0.35
\end{tabular}
\end{ruledtabular}
\end{table}

\section{Conclusions}
\label{sec:conclusions}
We have computed the total cross sections for $\gamma \gamma \rightarrow \phi\phi$, where $\phi \in \{h^0,A^0\}$, within the context of the Higgs Triplet model for a prospective muon collider with $\sqrt{s}=3 \ \mathrm{TeV}$. Our calculations take into account factors such as vacuum stability, the absence of tachyonic modes, unitarity conditions, and the $\rho$ parameter. Moreover, both $\lambda_4$ and the lepton flavor violation parameter play pivotal roles in our comprehensive calculations. For the numerical analysis, we defined two input parameter spaces: $\mathcal{P}_1$, generated by potential parameters ($\lambda$), and $\mathcal{P}_2$, generated by scalar boson masses and the vacuum expectation values of triplet and doublet fields in the Higgs triplet model. The evaluation of partonic cross sections involves the utilization of the $\mathcal{P}_2$ parameter space. It is observed that $\hat{\sigma}(\gamma\gamma \rightarrow h^0h^0)_{HTM} > \hat{\sigma}(\gamma\gamma \rightarrow h^0h^0)_{SM}$ for $\sqrt{\hat{s}}>450 \ \mathrm{GeV}$. 

In our analysis, we demonstrate that the cross sections for $\mu^+\mu^- \rightarrow \gamma \gamma \rightarrow h^0h^0, A^0A^0$ exhibit enhancement when $M_{H^0} \approx M_{A^0} > M_{H^{\pm}} > M_{H^{\pm\pm}}$, and the cross sections diminish to lower values when $M_{H^0} \approx M_{A^0}<M_{H^{\pm}} < M_{H^{\pm\pm}}$.
To perform these calculations, the $\mathcal{P}_1$ parameter space was employed, setting $\alpha \approx 0$ to ensure that the light \textit{CP}-even Higgs behaves similarly to the SM Higgs through three distinct approaches. Our first approach involved convoluting the partonic cross sections with the Weizsäcker-Williams approximation (LO), followed by the second approach, where we convoluted with the Weizsäcker-Williams approximation corrected up to the second order, $\mathcal{O}(\alpha_e^2t^2)$. Lastly, we employed the EW PDF set and discussed the differences among the results generated by each approach. Throughout our comprehensive calculations, the behavior of $\lambda_{hhh}^{HTM}$ closely aligns with $\lambda_{hhh}^{SM}$ numerically. Here, we explicitly demonstrate the significant contributions of $h^0H^{+}H^{-}$, $h^0h^0H^{+}H^{-}$, and $h^0A^0A^0$ in enhancing the cross sections of $\phi\phi$ production mechanisms. We have observed that the magnitude of the $H^0h^0h^0$ coupling, where $\sigma_{HTM} \sim \sigma_{SM}$, remains fixed for each hierarchy. We explored this using the aforementioned PDFs as well. These findings provide crucial insights into the comparable numerical behavior of the HTM and the SM regarding the $\lambda_{hhh}$ parameter.

\begin{acknowledgments}
We would like to express our sincere gratitude to the Bill Simons Foundation for Physics at Wichita State University for their generous funding during the summer of 2023. We also thank Wichita State University for supporting Bathiya by providing a unique applied learning experience working as a graduate assistant for the BeoShock High-Performance Computing Center. All numerical computations were performed using BeoShock High-Performance Computing at Wichita State University. A special thanks go to Keping Xie from Michigan State University for sharing the EW PDF set with us. 
\end{acknowledgments}

\appendix

\section{DECAY WIDTH OF $H^0$ AND $A^0$}
\label{app:decay}
\begin{figure}%\centering
\includegraphics[width=1.01\linewidth]{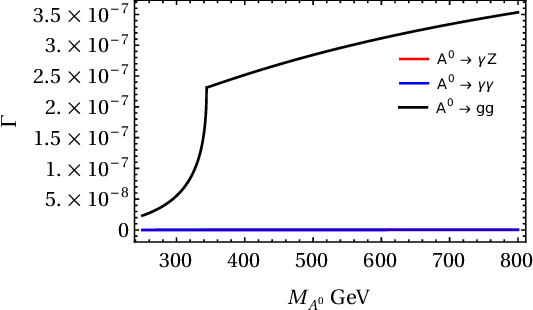} % Replace "figure.png" with the actual file name and extension
    \caption{The decay widths of $A^0\rightarrow\gamma\gamma, \gamma Z, gg$ as a function of $M_{A^0}$}\label{Decay1}
\end{figure}

\begin{figure}%\centering
\includegraphics[width=1.01\linewidth]{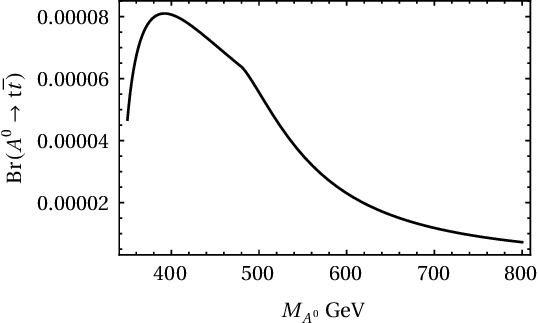} % Replace "figure.png" with the actual file name and extension
    \caption{The branching ratio of $A^0\rightarrow t\bar{t}$ as a function of $M_{A^0}$}\label{Br1}
\end{figure}

\begin{figure}%\centering
\includegraphics[width=1.01\linewidth]{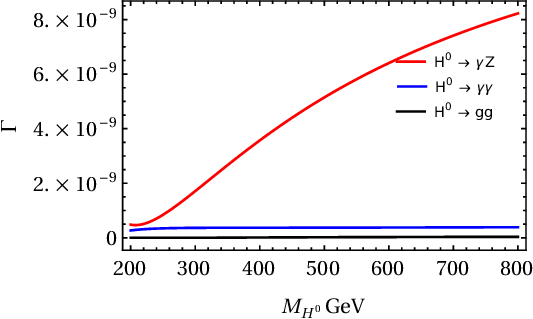} % Replace "figure.png" with the actual file name and extension
    \caption{The decay widths of $H^0\rightarrow \gamma Z, \gamma\gamma, gg$ as a function of $M_{H^0}~\mathrm{GeV}$. }\label{Br2}
\end{figure}

\begin{figure}%\centering
\includegraphics[width=1.01\linewidth]{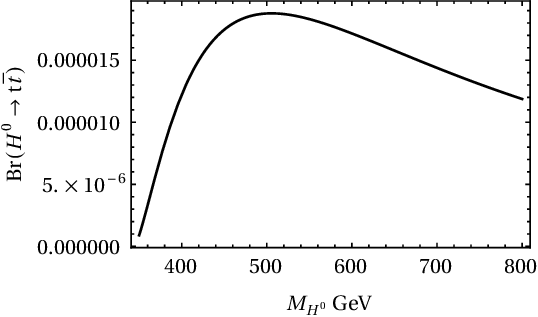} % Replace "figure.png" with the actual file name and extension
    \caption{The branching ratio of $H^0\rightarrow t\bar{t}$ as a function of $M_{H^0} \mathrm{GeV}$.}\label{Br3}
\end{figure}

\begin{figure}%\centering
\includegraphics[width=1.01\linewidth]{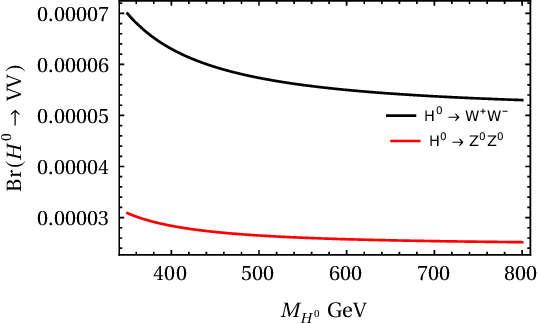} % Replace "figure.png" with the actual file name and extension
    \caption{The branching ratios of $H^0\rightarrow VV$, $V \in \{W^{\pm},Z^0\}$ as a function of $M_{H^0}~\mathrm{GeV}$.}\label{Br4}
\end{figure}

In this appendix, we provide some useful information about the decay widths and branching ratios of $H^0$ and $A^0$ that could be essential for further discussions. Figs~\ref{Decay1}- \ref{Br4} show the numerical results of the decay widths and branching ratios according to the parameter choices in Section \ref{sec:results}. The branching ratios of $H^0 \rightarrow t\bar{t},W^+W^-, Z^0Z^0$, and $A^0 \rightarrow t\bar{t}$ were calculated using {\tt FeynRules} \cite{Christensen:2008py}.

The decay channels of $A^0/H^0 \rightarrow \gamma\gamma, \gamma Z^0, gg$ are given by analytic expressions \ref{eq:deacay1} and \ref{eq:deacay2}. The contributions from $H^{\pm}$ and $H^{\pm\pm}$ should be considered to calculate the decay widths of $H^0 \rightarrow \gamma\gamma, \gamma Z^0$ in the context of HTM. 

\begin{equation}\label{eq:deacay1}
\Gamma(A^0/H^0\rightarrow \gamma\gamma,gg) = \frac{|\mathcal{M}(A^0/H^0\rightarrow \gamma\gamma,gg)|^2}{32\pi M_{A^0/H^0}},
\end{equation}

\begin{equation}\label{eq:deacay2}
\Gamma(A^0/H^0\rightarrow \gamma Z^0) = \frac{|\mathcal{M}(A^0/H^0\rightarrow \gamma Z^0)|^2}{16\pi M_{A^0/H^0}}\Big(1-\frac{M_{Z}^2}{M_{A^0/H^0}^2}\Big).
\end{equation}
The analytic expression for $|\mathcal{M}|$ is given by,
\begin{align}\label{eq:Msqrd}
|\mathcal{M}(A^0/H^0\rightarrow XX)|^2 &= |\mathcal{M}_{++}(A^0/H^0\rightarrow XX)|^2 \nonumber\\
&\quad + |\mathcal{M}_{--}(A^0/H^0\rightarrow XX)|^2,
\end{align}
    where $XX = \gamma\gamma, gg, \gamma Z^0$. 
    According to \texttt{FormCalc} output, we were able to confirm that $\mathcal{M}_{+-} = \mathcal{M}_{-+} = 0$ and $|\mathcal{M}_{+-}|^2 = |\mathcal{M}_{-+}|^2$. The analytic expressions for $\mathcal{M}_{++}$ are shown in \ref{eq:Mppaa}-\ref{eq:MPPaZ}. In these expressions, $f$ and 
$q$ represent fermions and quarks, respectively, where 
 $N_f^c=3$ for quarks and $1$ for leptons. $Q_f$ denotes the charge of the Standard Model fermions. The following scalar functions used in these expressions are evaluated with {\tt LoopTools}, Ref.~\cite{Hahn:1998yk}.
\begin{equation}\label{loopIn1}
    B_0(p^2,m_1^2,m_2^2)=\int\frac{d^Dq}{i\pi^2}\frac{1}{(q^2-m_1^2)((q+p)^2-m_2^2)},
\end{equation}
\begin{align}
 C_0,C^{\mu},C^{\mu \nu}(p_1^2,p_2^2,(p_1+p_2)^2,m_1^2,m_2^2,m_3^2) &=\nonumber\\
    \int\frac{d^Dq}{i\pi}\frac{1,q^{\mu},q^{\mu}q^{\nu}}{(q^2-m_1^2)((q+p_1)^2-m_2^2)((q+p_1+p_2)^2-m_3^2)}.
\end{align}
    
The decay rates for the Higgs boson decaying into two gauge bosons, $VV$ $(V = W^{\pm}, Z^0)$, are given by, Ref.~\cite{Arbabifar:2012bd}:

\begin{equation}
\Gamma(H^0\rightarrow VV) = \frac{\delta_V}{512 \pi} \eta_V^2 M_H^3 \left[ 1-\frac{4M_V^2}{M_H^2}+\frac{12M^2_Z}{M_H^4}\right]\beta\left(\frac{M^2_V}{M_H^2}\right),
\end{equation}

where $\delta=2$ for $W^{\pm}$ and $\delta=1$ for $Z^0$, and

\begin{equation}
    \eta_W = \frac{g^2}{M_W^2}(s_{\alpha}v_{\Phi}-2c_{\alpha}v_{\Delta}),
\end{equation}

\begin{equation}
    \eta_Z = \frac{g^2}{M_Z^2C_W^2}(s_{\alpha}v_{\Phi}-4c_{\alpha}v_{\Delta}).
\end{equation}

The decay rate for the Higgs boson decaying into two fermions is given by:

\begin{equation}
    \Gamma(H^0 \rightarrow f\bar{f}) = \frac{g^2}{32\pi M_W^2} M_H^2 M_f^2 N^c_f \beta\left(\frac{M_f^2}{M_H^2}\right)^3\sin^2{\alpha},
\end{equation}
where $M_f$ is the fermion mass.
Note that,
\begin{equation}\label{eq:beta}
    \beta\left(\frac{M_{V,f}^2}{M_H^2}\right) = \sqrt{1-\frac{4M^2_{V,f}}{M_H^2}}.
\end{equation}
%The invisible decay width is given by:
%\begin{equation}
%    \Gamma(H^0 \rightarrow \nu \nu)= \sum_{i,j=1}^{3}S_{ij}|h_{ij}|^2\frac{M_H}{4\pi}\cos^2{\alpha},
%\end{equation}
%where $S_{ij}$ and $h_{ij}$ are specific parameters related to the neutrino mixing, Ref.~\cite{Arbabifar:2012bd}. 
\newpage
\begin{widetext}
\begin{align}
    \mathcal{M_{++}}(A^0\rightarrow \gamma\gamma) &= \frac{i}{\sqrt{2}\pi}\alpha_e M_{A^0}^2\sin{\beta}\sum_{f}N^c_f \mathrm{sgn}(Q_f)Q_f^2M_fy_fC_{0}(0,0,M_{A^0},M_{f}^2,M_{f}^2,M_{f}^2)\label{eq:Mppaa}, \\
    \mathcal{M_{++}}(A^0\rightarrow gg) &= \frac{i}{\sqrt{2}\pi}\alpha_sM_{A^0}^2\sin{\beta}\sum_{q}\mathrm{sgn}(Q_q)M_qy_qC_{0}(0,0,M_{A^0},M_{f}^2,M_{f}^2,M_{f}^2)\delta^{ab},
    \\
    \mathcal{M_{++}}(A^0\rightarrow \gamma Z^0) &=\frac{\alpha_e \sin{\beta}M_{A^0}^2}{4\sqrt{2}\pi C_W S_W}\Big(1-\frac{M_Z^2}{M_{A^0}^2}\Big)\sum_fQ_fN_f^cM_fy_f(1-4Q_fS_W^2)C_{0}(M_Z^2,0,M_{A^0},M_{f}^2,M_{f}^2,M_{f}^2),
    \\
    \mathcal{M_{++}}(H^0\rightarrow gg) &= \frac{\alpha_s}{\sqrt{2}\pi}\sin{\alpha}\sum_q M_qy_qF^{gg}_q \delta^{ab},
    \\
    \mathcal{M_{++}}(H^0\rightarrow \gamma\gamma) &= \mathcal{M_{++}^{\mathrm{f}}}(H^0\rightarrow \gamma\gamma)+\mathcal{M_{++}^{\mathrm{W}}}(H^0\rightarrow \gamma\gamma)+\mathcal{M_{++}^{\mathrm{H^{\pm}}}}(H^0\rightarrow \gamma\gamma)+\mathcal{M_{++}^{\mathrm{H^{\pm\pm}}}}(H^0\rightarrow \gamma\gamma),
    \\
    \mathcal{M_{++}^{\mathrm{f}}}(H^0\rightarrow \gamma\gamma) &= \frac{\alpha_e}{\sqrt{2}\pi}\sum_fN^c_fQ_f^2M_fy_fF^{\gamma\gamma}_f,
    \\
    \mathcal{M_{++}^{\mathrm{W}}}(H^0\rightarrow \gamma\gamma) &= \frac{\alpha_e}{2S_W^2}(\sin{\alpha}v_{\Phi}-2\cos{\alpha}v_{\Delta})F_W^{\gamma\gamma},
    \\
    \mathcal{M_{++}^{\mathrm{H^{\pm}}}}(H^0\rightarrow \gamma\gamma) &= \frac{\alpha_e}{\pi}\lambda^{HTM}_{H^0H^{\pm}H^{\pm}}C_0(0,0,M_{H^0}^2,M_{H^{\pm}}^2,M_{H^{\pm}}^2,M_{H^{\pm}}^2),
    \\
    \mathcal{M_{++}^{\mathrm{H^{\pm\pm}}}}(H^0\rightarrow \gamma\gamma) &= \frac{4\alpha_e}{\pi}\lambda^{HTM}_{H^0H^{\pm\pm}H^{\pm\pm}}C_0(0,0,M_{H^0}^2,M_{H^{\pm\pm}}^2,M_{H^{\pm\pm}}^2,M_{H^{\pm\pm}}^2),
    \\
    \mathcal{M_{++}}(H^0\rightarrow \gamma Z^0) &= \mathcal{M_{++}^{\mathrm{f}}}(H^0\rightarrow \gamma Z^0)+\mathcal{M_{++}^{\mathrm{W}}}(H^0\rightarrow \gamma Z^0)+\mathcal{M_{++}^{\mathrm{H^{\pm}}}}(H^0\rightarrow \gamma Z^0)+\mathcal{M_{++}^{\mathrm{H^{\pm\pm}}}}(H^0\rightarrow \gamma Z^0),
    \\
   \mathcal{M_{++}^{\mathrm{f}}}(H^0\rightarrow \gamma Z^0) &= \frac{\alpha_e\sin{\beta}}{4\sqrt{2}\pi C_W S_W} \sum_f |Q_f| N^c_f M_fy_f(1-4|Q_f|S_W^2)F_f^{\gamma Z},
   \\
   \mathcal{M_{++}^{\mathrm{H^{\pm}}}}(H^0\rightarrow \gamma Z^0) &= -\frac{\alpha_e}{4\pi S_W C_W}\lambda^{HTM}_{H^0H^{\pm}H^{\pm}}(S_W^2-C_W^2+\cos{\beta^{'}})F^{\gamma Z}_{H^{\pm}},
   \\
\mathcal{M_{++}^{\mathrm{H^{\pm\pm}}}}(H^0\rightarrow \gamma Z^0) &= \frac{\alpha_e}{\pi C_W S_W}(C_W^2-S_W^2)\lambda^{HTM}_{H^0H^{\pm\pm}H^{\pm\pm}}F^{\gamma Z}_{H^{\pm\pm}},
\\
\mathcal{M}_{++}^{\mathrm{W}}(H^0\rightarrow \gamma Z^0) &= \eta_1B_0(0,M_W^2,M_W^2)+\eta_2B_0(M_{H^0}^2,M_W^2,M_W^2)+\eta_3B_0(M_Z^2,M_W^2,M_W^2)\label{eq:MPPaZ}\\
&\quad+\eta_4C_0(0,M_Z^2,M_{H^0}^2,M_W^2,M_W^2,M_{H^{\pm}}^2)+\eta_{5}C_0(0,M_Z^2,M_{H^0}^2,M_W^2,M_W^2,M_W^2)\nonumber\\
&\quad+\eta_6C_1(0,M_Z^2,M_{H^0}^2,M_W^2,M_W^2,M_W^2)+\eta_7C_1(0,M_Z^2,M_{H^0}^2,M_W^2,M_W^2,M_{H^{\pm}}^2)\nonumber\\
&\quad+\eta_8C_2(0,M_Z^2,M_{H^0}^2,M_W^2,M_W^2,M_{H^{\pm}}^2)+\eta_9C_2(0,M_Z^2,M_{H^0}^2,M_W^2,M_W^2,M_W^2)\nonumber\\
&\quad+\eta_{10}C_{00}(0,M_Z^2,M_{H^0}^2,M_W^2,M_W^2,M_W^2)+\eta_{11}C_{00}(M_{H^0}^2,0,M_Z^2,M_W^2,M_{H^{\pm}}^2,M_{H^{\pm}}^2)\nonumber\\
&\quad+\eta_{12}C_{00}(0,M_Z^2,M_W^2,M_W^2,M_{H^{\pm}}^2)\nonumber,\\
\eta_1&=\frac{2\alpha_ev_{\Delta}^3(c_{\alpha}v_{\Phi}+s_{\alpha}v_{\Delta})}{C_WS_W^3v_{\Phi}(2v^2_{\Phi}+v_{\Delta}^2)},\\
\eta_2&=\frac{3\alpha_eC_W(s_{\alpha}v_{\Phi}-2c_{\alpha}v_{\Delta})}{S_W^3},\\
\eta_3&=\frac{\alpha_e^2\left(s_\alpha v_\phi-2 c_\alpha v_{\Delta}\right)}{2 C_W S_W^2 v_\phi^2\left(2 v_\phi^2+2 v_{\Delta}^2\right)}\left\{8 C_W^2 v_\phi^4+\left(S_W^2-3\right) v_\phi^2 v_{\Delta}^2 -2\left(1+3 S_W^2\right) v_{\Delta}^4\right\},
\\
\eta_4&=-\frac{2 \alpha_e v_{\Delta}^3\left(c_\alpha v_\phi+s_a v_{\Delta}\right)\left(v_\phi^4+v_{\Delta}^2 v_\phi^2+v_{\Delta}^4\right)}{C_W S_W^3 v_\phi^2\left(2 v_\phi^2+v_{\Delta}^2\right)^2},
\end{align}
%\end{widetext}
%\begin{widetext}
\begin{align}
\eta_5 &= \frac{\alpha_e^2}{4 C_W S_W^5 v_\phi^3\left(2 v_\phi^2+v_{\Delta}^2\right)^2} \times \bigg\{S_W^2 v_\phi\left(s_\alpha v_\phi-2 c_\alpha v_{\Delta}\right)\left(2 v_\phi^2+v_{\Delta}^2\right)\nonumber \\
& \quad \times \bigg(16\left(M_{H^0}^2-M_{W}^2 - M_Z^2\right) C_W^2 v_\phi^4+\left(-7 M_{H^0}^2+6 M_W^2+7 M_Z^2+\left(11 M_{H^0}^2-4 M_W^2-11 M_Z\right) S_W^2\right) \nonumber \\
& \quad \times v_\phi^2 v_{\Delta}^2 +2\left(5 M_{H^0}^2-5 M_Z^2+\left(3 M_{H^0}^2+4 M_W^2-3 M_Z^2\right) S_W^2\right) v_{\Delta}^4 -v_{\Delta}^2\left(v_\phi^2+2 v_{\Delta}^2\right)\nonumber \\
& \quad \times \bigg(-\left(\left(M_{H^0}^2+M_Z^2\right) S_W^2 C_W^2 v_\phi\left(s_\alpha v_\phi-2 c_\alpha v_{\Delta}\right)\right.\left.\left(2 v_\phi^2+v_{\Delta}^2\right)\right)+\left(S_W^2 v_\phi^2+2\left(1+S_W^2\right) v_{\Delta}^2\right) \nonumber \\
& \quad \times \bigg(-4 \alpha_e \pi v_\phi\left(s_\alpha v_\phi-2 c_\alpha v_{\Delta}\right)\left(2 v_\phi^2+v_{\Delta}^2\right)+S_W^2\left(-2\left(2 \lambda_1+\lambda_4\right) s_\alpha v_\phi^4+8 c_\alpha\left(\lambda_2+\lambda_3\right) v_\phi^3 v_{\Delta} \right. \nonumber \\
& \quad -\lambda s_\alpha v_\phi^2 v_{\Delta}^2+2\left(2 \sqrt{2} \mu s_\alpha+c_\alpha\left(\lambda_1+\lambda_4\right) v_\phi\right) v_{\Delta}^3 -2 \lambda_4 c_\alpha v_{\Delta}^4\bigg)\bigg)\bigg\},
\\
\eta_6&=\frac{\alpha_e^2\left(M_{H^0}^2-M_Z^2\right)\left(S_\alpha v_\phi-2 C_\alpha v_{\Delta}\right)}{2 C_W S_W^3\left(2 V_\phi^2+v_{\Delta}^2\right)}\left\{-2 C_W^2 v_\phi^4-\left(1+C_W^2\right) v_{\Delta}^2 v_\phi^2-4 v_{\Delta}^4 \right\},
\\
\eta_7&=\frac{\alpha_e^2}{C_W S_W^3 v_\phi\left(2 v_\phi^2+v_{\Delta}^2\right)}\left(M_{H^0}^2- M_Z^2\right) v_{\Delta}^3\left(c_\alpha v_\phi+s_\alpha v_{\Delta}\right), \\
\eta_8&=\frac{\alpha_e^2}{C_W S_W^3}\left(3 M_{H^0}^2-M_Z^2\right) v_{\Delta}^3\left(c_\alpha v_\phi+s_{\alpha} v_{\Delta}\right),
\\
  \eta_9 &= \frac{\alpha_e^2\left(s_{\alpha}v_\phi-2 c_\alpha v_{\Delta}\right)}{2 c_W S_W^3 v_\phi^2\left(2 v_\phi^2+v_{\Delta}^2\right)}\left\{4\left(2 M_{H^0}^2-M_Z^2\right) C_W^2 v_\phi^4+\left(M_Z^2+M_{H^0}^2\left(2 S_W^2-5\right)\right) v_\phi^2 v_{\Delta}^2 \right. \nonumber \\
  &\quad -4\left(2 M_{H^0}^2+\left(M_{H^0}^2-M_{Z}^2\right) S_W^2\right) v_{\Delta}^4\left. \right\}, 
  \\
  \eta_{10}&=\frac{\alpha_e^2\left(s_\alpha v_\phi-2 c_\alpha v_{\Delta}\right)}{C_W S_W^3 v_\phi^2\left(2 v_\phi^2+v_{\Delta}^2\right)}\left\{40 C_W^2 v_\phi^6+\left(-41+46 S_W^2\right) v_\phi^4 v_{\Delta}^2
+\left(-7+25 S_W^2\right) v_\phi^2 v_{\Delta}^4+\left(1+6 S_W^2\right) v_{\Delta}^6\right\},
\\
\eta_{11}&=\frac{-2 \alpha_e^2 v_{\Delta}^3\left(c_\alpha v_\phi+S_\alpha v_{\Delta}\right)}{C_W S_W^3 v_\phi\left(2 v_\phi^2+v_{\Delta}^2\right)},
\\
\eta_{12}&= \frac{\alpha_e^2 v_{\Delta}^3}{4 C_W S_W^3\left(2 v_\phi^2+v_{\Delta}^2\right)^2}\bigg\{4(2M_{H^0}^{2} + M_{W}^{2} - 2M_{Z}^{2})v_{\phi}(c_{\alpha}v_{\phi} + s_{\alpha} v_{\Delta})\left(2 v_\phi^2+v_{\Delta}^2\right) - \left(v_\phi^2 + 2 v_{\Delta}^2\right)\bigg(-2 \lambda_4 s_\alpha v_\phi^3 v_{\Delta} \nonumber \\
& \quad + \left(-2 \lambda^2 + 4 \lambda_1 + 3 \lambda_4\right) s_\alpha v_\phi v_{\Delta}^3 + 2 \sqrt{2} \mu s_\alpha v_\phi\left(2 v_\phi^2 - v_{\Delta}^2\right) + c_\alpha\left(2 \lambda_4 v_\phi^4 - \lambda_4 v_\phi^2 v_{\Delta}^2 + 4\left(\lambda_1 - 2\left(\lambda_2 + \lambda_3\right)\right) v_{\Delta}^4\right)\bigg)\bigg\},
\\
 F_W^{\gamma \gamma}&=4 B_0\left(0, M_W^2, M_W^2\right)-6 B_0\left(M_{H^0}^2, M_W^2, M_W^2\right)+4\left(M_W^2-M_{H^0}^2\right) C_0\left(0,0, M_{H^0}^2, M_W^2, M_W^2, M_W^2\right)\nonumber\\
& +20 C_{00}\left(0,0, M_{H^0}^2, M_W^2, M_W^2, M_W^2\right)+M_{H^{0}}^2 \left\{ C_1\left(0,0, M_{H^{0}}^2, M_W^2, M_W^2, M_W^2\right)+4C_2\left(0,0, M_{H^0}^2, M_W^2, M_W^2, M_W^2\right)\right\},
\\
F_{f,q}^{\gamma\gamma,gg}&=2 B_0\left(0, M_{f,g}^2, M_{f,g}^2\right)+M_{H^0}^2 C_0\left(0,0, M_{H^0}^2, M_{f,g}^2, M_{f,g}^2, M_{f,g}^2\right)-8 C_{00}\left(0,0, M_{H^0}^2, M_{f,g}^2, M_{f,g}^2, M_{f,g}^2\right)\nonumber \\&+2 C_2\left(0,0, M_{H^0}^2, M_{f,g}^2, M_{f,g}^2, M_{f,g}^2\right),
\\
F_{H^{\pm}}^{\gamma Z}&=B_0\left(M_{H^0}^2, M_{H^{\pm}}^2, M_{H^{\pm}}^2\right)-4 C_{00}\left(0, M_Z^2, M_{H^0}^2, M_{H^{\pm}}^2, M_{H^{\pm}}^2, M_{H^{\pm}}^2\right),
\\
F_{H^{ \pm \pm}}^{\gamma Z}&=B_0\left(M_{H^0}^2, M_{H^{ \pm \pm}}^2, M_{H^{ \pm \pm}}^2\right)-4 C_{00}\left(0, M_Z^2, M_{H^0}^2, M_{H^{\pm\pm}}^2, M_{H^{\pm\pm}}^2 M_{H^{\pm\pm}}^2\right).
\end{align}
\end{widetext}

\section{}\label{sec:AppB}
In order to ensure that $M_H^2 \geq 0$, we examine the function $f(\mu)= -a\mu^2+b\mu+c$, which must be non-negative. To have a maximum value for $f(\mu)$, we require $a > 0$. Here, the coefficients are given by:

\begin{equation}
\begin{aligned}
    & a = 8\pi\alpha_{e}\left(1+\frac{\tan^2\beta}{2}\right)\tan{\beta}\\
    & b = 4\sqrt{\pi\alpha_{e}}M_WS_W \left(1+\frac{\tan^2\beta}{2}\right)\{ \lambda+2(\lambda_1+\lambda_3)\tan{\beta}\} \\
    & c = -((\lambda_1+\lambda_4)^2-\lambda(\lambda_2+\lambda_3))M_W^2S^2_W\tan^3{\beta}.
\end{aligned}
\end{equation}

In Eq.~(\ref{eq:widemu}), the expression inside the square brackets can be written as:

\begin{equation}
\begin{aligned}
2\lambda^2+8\lambda\tan^2{\beta}(\lambda_1+\lambda_4+\lambda_2\tan^2{\beta}+\lambda_3\tan^2{\beta}) \\ \geq 2\lambda(\sqrt{\lambda}-2\tan^2{\beta}\sqrt{\lambda_2+\lambda_3})^2 \geq 0,
\end{aligned}   
\end{equation}

under the BFB (bounded from below) conditions mentioned in Sec.~\ref{sec:potential}. Therefore, the roots of $f(\mu)=0$, $\mu_{\pm} \in \mathbb{R}$. 

\section{}

The following expression provides the second order corrected EPA that we use in implementing PDFs using a Fortran subroutine. Further details are provided in Sec.~2 of Ref.~\cite{Garosi:2023bvq}.
\begin{equation}\label{LLf}
    f_{\gamma}(x,t) = \frac{\alpha_e}{2\pi}tP_{vf}^{f}+\frac{1}{2}\Big(\frac{\alpha_e t}{2\pi}\Big)^2[(P^v_{\gamma}+P^v_f)P^f_{vf}+I_{vfff}]
\end{equation}
Where,
\begin{equation}\nonumber
    t = \alpha_e \log\Big(\frac{E^2}{M_{\mu}}\Big),
    P_f^v = \frac{3}{2},
    P^v_f = \frac{-40}{9},
    P_{vf}^f = \frac{1+(1-x)^2}{x}
\end{equation}
\begin{equation}\nonumber
    I_{vfff}= \Big( \frac{3}{2}+2\log(1-x)\Big), 
\end{equation}
\begin{equation}\nonumber
   P^f_{vf}=\frac{(1-x)(2x-3)}{x}+(2-x)\log(x).
\end{equation}

\section{}
To maintain numerical stability when implementing the Fortran subroutines, we employed the transformations $x=\xi^2_1+(1-\xi^2_1)\xi_2$ and $\tau=\xi^2_1$. The Eq.~(\ref{fullXsec}) has been modified as follows.
\begin{align}
\sigma(\mu^+\mu^- &\rightarrow \gamma\gamma \rightarrow \phi\phi)\nonumber \\
&= 2 \int_{\sqrt{\tau_{m}}}^{1} \int_{0}^{1} \frac{(1-\xi_1^2)}{[\xi_1^2+(1-\xi_1^2)\xi_2]}\frac{\hat{F}}{\xi_1}d\xi_1d\xi_2\hat{\sigma}(\hat{s})\big|_{\hat{s}=\xi_1^2s} 
\end{align}

\begin{align}
\hat{F}= xf_{\gamma}(x)\frac{\tau}{x}f_{\gamma}(\frac{\tau}{x})\big|_{x=\xi^2_1+(1-\xi^2_1)\xi_2,\tau=\xi^2_1}
\end{align}

\section{FEYNMAN RULES FOR 3 AND 4 SCALARS}\label{FeynRules}
We have provided below a set of Feynman rules applicable for any value of the mixing angle ($\alpha$) between the doublet and triplet fields, see the Table~\ref{tab4}. 
\begin{table*}
\caption{\label{tab4}Here, we adopt the notation $c_{\beta}\equiv \cos{\beta}$, $s_{\beta}\equiv \sin{\beta}$, $c_{\beta^{'}}\equiv \cos{\beta^{'}}$, and $s_{\beta^{'}}\equiv \sin{\beta^{'}}$.}
\begin{ruledtabular}
\begin{tabular}{c|c}
   Interaction & Feynman Rule   \\
  \hline $h^0h^0h^0$  & 
 $ -6i \Bigl\{ -\frac{\mu s_{\alpha}c_{\alpha}^2}{\sqrt{2}}+\frac{c_{\alpha}v_{\phi}}{2}\Big(\frac{\lambda c_{\alpha}^2}{2}+(\lambda_1+\lambda_4)s_{\alpha}^2\Big)+\frac{s_{\alpha}v_{\Delta}}{2}\Big({\lambda_1+\lambda_4) c_{\alpha}^2}+2(\lambda_2+\lambda_3)s_{\alpha}^2\Big)
 \Bigr\}$  \\
   $h^0h^0H^0$      & $ -\frac{i}{2} \Bigl\{-2\sqrt{2}\mu c_{\alpha}(1-3c_{\alpha}^2)-2\Big(\frac{3\lambda}{2}c_{\alpha}^2 + (\lambda_1+\lambda_4)(1-3c_{\alpha}^2)\Big)s_{\alpha}v_{\phi} +2\Big(6(\lambda_2+\lambda_4)s_{\alpha}^2+(\lambda_1+\lambda_4)(1-3c_{\alpha}^2)\Big)c_{\alpha}v_{\Delta}\Bigr\}$  \\
   $h^0A^0A^0$  & $  -\frac{i}{2}\Big\{2\sqrt{2}\mu s_{\beta}(2c_{\alpha}c_{\beta}+s_{\alpha}s_{\beta})+ (\lambda s^2_{\beta}+2(\lambda_1+\lambda_4)c^2_{\beta})v_{\phi}c_{\alpha}+2(2(\lambda_2+\lambda_3)c^2_{\beta}+(\lambda_1+\lambda_4)s^2_{\beta}))v_{\Delta}s_{\alpha}\Big\}$ \\ $H^0A^0A^0$ &$  -\frac{i}{2}\Big\{-2\sqrt{2}\mu s_{\beta}(2s_{\alpha}c_{\beta}-c_{\alpha}s_{\beta})- (\lambda s^2_{\beta}+2(\lambda_1+\lambda_4)c^2_{\beta})v_{\phi}s_{\alpha}+2(2(\lambda_2+\lambda_3)c^2_{\beta}+(\lambda_1+\lambda_4)s^2_{\beta}))v_{\Delta}s_{\alpha}\Big\}$\\
   $h^0H^{+}H^{-}$&$-\frac{i}{2}\Big\{ (\lambda v_{\phi}s^2_{\beta^{'}}+2\lambda_1v_{\phi}c^2_{\beta^{'}}+\lambda_4v_{\phi}c_{2\beta^{'}}+2\mu s_{2\beta^{'}})c_{\alpha}+(4(\lambda_2+\lambda_3)v_{\Delta}c^2_{\beta^{'}}+2\lambda_1v_{\Delta}s^2_{\beta^{'}}-2v_{\Delta}\lambda_4c^2_{\beta^{'}})s_{\alpha}\Big\}$\\
    $H^0H^{+}H^{-}$&$-\frac{i}{2}\Big\{ -(\lambda v_{\phi}s^2_{\beta^{'}}+2\lambda_1v_{\phi}c^2_{\beta^{'}}+\lambda_4v_{\phi}c_{2\beta^{'}}+2\mu s_{2\beta^{'}})s_{\alpha}+(4(\lambda_2+\lambda_3)v_{\Delta}c^2_{\beta^{'}}+2\lambda_1v_{\Delta}s^2_{\beta^{'}}-2v_{\Delta}\lambda_4c^2_{\beta^{'}})c_{\alpha}\Big\}$\\
   $h^0H^{++}H^{--}$&$ -ic_{\alpha}\lambda_1v_{\phi}-2i\lambda_2 s_{\alpha}v_{\Delta}$\\
   $H^0H^{++}H^{--}$&$ is_{\alpha}\lambda_1v_{\phi}-2i\lambda_2 c_{\alpha}v_{\Delta}$\\
 $h^0h^0H^{+}H^{-}$&$ -\frac{i}{2}\Big\{ \lambda c^{2}_{\alpha}s^2_{\beta^{'}}+2\lambda_1(c_{\beta^{'}}^2 c^2_{\alpha}+ s_{\beta^{'}}^2s^2_{\alpha}) + 4c_{\beta^{'}}^2s_{\alpha}^2(\lambda_2+\lambda_3)-\lambda_4\Bigl(c_{\beta^{'}}^2s_{\alpha}^2+4c^2_{\beta^{'}}s_{\alpha}c_{\alpha}\frac{v_{\Delta}}{v_{\Phi}}\Bigr)\Big\}$\\
 $A^0A^0H^{+}H^{-}$& $-\frac{i}{2}\Big\{ -\lambda(1+c^2_{\beta}+c^2_{\beta^{'}}s^2_{\beta})+2\lambda_1(c_{\beta}+c_{\beta^{'}})^2+4(c^2_{\beta}c^2_{\beta^{'}}(\lambda_2+\lambda_3)+\lambda_4\Bigl(c_{\beta^{'}}^2c_{\alpha}^2+4\sqrt{2}c^2_{\beta^{'}}s_{\beta^{'}}c_{\alpha}\frac{v_{\Delta}}{v_{\Phi}}\Bigr)\Big\}$\\
$ h^0h^0H^{++}H^{--}$&$-i\lambda_1+i\lambda_1s_{\alpha}^2-2i\lambda_2s^2_{\alpha}$\\
$A^0A^0H^{++}H^{--}$&$-i\lambda_1+i\lambda_1c_{\beta}^2-2i\lambda_2c^2_{\beta}$

   \end{tabular}
\end{ruledtabular}
\end{table*}

\section{$\tan \beta$ EXPANDED COUPLING FACTORS }
\label{app:series}
In this appendix we present the expanded form of the coupling factor $\lambda^{HTM}$ for $3$ and $4$ point scalar interactions. 
\begin{equation}\label{cup1}
        \lambda_{h^0h^0h^0}^{HTM}= \frac{3}{2}\frac{M_WS_W}{\sqrt{\pi\alpha_{e}}}\lambda - \frac{3}{8}\lambda \frac{M_WS_W}{\sqrt{\pi\alpha_{e}}}\tan^2{\beta} + \mathcal{O}(\tan^4{\beta})
\end{equation}

\begin{align}\label{cup2}
\lambda^{HTM}_{H^0h^0h^0}=\sqrt{2}\mu-\frac{(\lambda_1+\lambda_4)M_WS_W}{2\sqrt{\alpha_{e}\pi}}\tan{\beta}\nonumber\\+ \frac{(\lambda_1+\lambda_4)M_WS_W}{8\sqrt{\alpha_{e}\pi}}\tan^3{\beta}+ \mathcal{O}(\tan^5{\beta})
\end{align}
\begin{align}\label{cup3}
\lambda^{HTM}_{H^0A^0A^0}= \frac{(\lambda_2+\lambda_3)M_WS_W}{\sqrt{\alpha_{e}\pi}}\tan{\beta} + \sqrt{2}\mu\tan^2{\beta}\nonumber\\+\frac{(2\lambda_1-5\lambda_2-5\lambda_3+2\lambda_4)M_WS_W}{4\sqrt{\alpha_{e}\pi}}\tan^3{\beta}+\mathcal{O}(\tan^4{\beta})
\end{align}
\begin{align}\label{cup4}
    \lambda_{h^0H^{+}H^{-}}^{HTM} = -\frac{(2\lambda_1+\lambda_4)M_WS_W}{2\sqrt{\alpha_e\pi}}-\sqrt{2}\mu\tan{\beta}\nonumber\\ -\frac{(2\lambda-6\lambda_1-5\lambda_4)M_WS_W}{8\sqrt{\alpha_e\pi}} \tan^2{\beta}+\mathcal{O}(\tan^4{\beta})
\end{align}
\begin{align}\label{cup5}
    \lambda_{h^0H^{++}H^{--}}^{HTM} \approx -\frac{\lambda_1M_WS_W}{\sqrt{\alpha_e\pi}}
\end{align}
\begin{align}\label{cup6}
    \lambda_{h^0h^0H^{+}H^{-}}^{HTM} = -\frac{1}{2}(\lambda_1+\lambda_4)+\frac{1}{4}(-\lambda+2\lambda_1+\lambda_4)\tan^2{\beta}\nonumber\\ + \frac{1}{8}(\lambda-2\lambda_1-\lambda_4)\tan^4{\beta}+\mathcal{O}(\tan^6{\beta})
\end{align}
\begin{align}\label{cup7}
    \lambda_{h^0h^0H^{++}H^{--}}^{HTM} \approx \lambda_1
\end{align}
\begin{align}\label{cup8}
    \lambda_{h^0A^0A^0}^{HTM} = -\frac{(\lambda_1+\lambda_4)M_WS_W}{\sqrt{\alpha_e\pi}}-2\sqrt{2}\mu\tan{\beta}\nonumber\\+\frac{(-2\lambda+5\lambda_1+5\lambda_4)M_WS_W}{8\sqrt{\alpha_e\pi}}\tan^2{\beta}+\mathcal{O}(\tan^3{\beta})
\end{align}
\begin{align}\label{cup9}
    \lambda_{A^0A^0H^{++}H^{--}}^{HTM} = -2\lambda_2+(-\lambda_1+2\lambda_2)\tan^2{\beta}+\mathcal{O}(\tan^4{\beta})
\end{align}
\begin{align}\label{cup10}
    \lambda_{A^0A^0H^{+}H^{-}}^{HTM} = -2(\lambda_2+\lambda_3)\nonumber\\+
    \frac{1}{2}(6(\lambda_2+\lambda_3)-3(\lambda_1+\lambda_4))\tan^2{\beta}+\mathcal{O}(\tan^4{\beta})
\end{align}

\bibliography{main}

\end{document}